\documentclass[aps,twocolumn,floats]{revtex4-2}
\usepackage[margin=0.8in,footskip=0.25in,paperwidth=8in,paperheight=13.5in]{geometry}
\usepackage{latexsym}
\usepackage{balance}
\usepackage{marginnote}
\usepackage{lipsum}
\usepackage{amsmath}
\usepackage{amsmath}

\usepackage{amssymb}
\usepackage{bm}
\usepackage{wasysym}
\usepackage[dvips]{color}
\usepackage{graphicx}
\usepackage{subfigure}
\usepackage{epsfig}
\usepackage{balance}
\usepackage{subfigure}
\usepackage{ulem}

\DeclareMathAlphabet{\mathpzc}{OT1}{pzc}{m}{it}

\newcommand{\hide}[1]{}
\newcommand{\veps}{\varepsilon}

\def\bfp{{\bf p}}
\def\bfr{{\bf r}}

\def\veps{\varepsilon}
\def\ua{\uparrow}
\def\da{\downarrow}

\usepackage{tikz}
\usetikzlibrary{matrix}
\usetikzlibrary{arrows}

\graphicspath{{./figures/}}
\usepackage{float}

\usepackage[colorlinks,bookmarks=true,citecolor=blue,linkcolor=blue,urlcolor=blue, breaklinks=true]{hyperref}
\usepackage[sort&compress]{natbib}

\newcommand{\be}{\mathrm{e}}

\begin{document}
\title{Aharonov--Bohm and Aharonov--Casher effects in meso-scopic physics: A brief review}
\author{Y. Avishai$^{1,2}$, Y. B. Band$^{1,3,4}$}
\affiliation{$^1$Department of Physics,
  Ben-Gurion University of the Negev,
  Beer-Sheva, Israel, \\
  $^2$Yukawa Institute for Theoretical Physics, Kyoto, Japan\\
  $^3$Department of Chemistry,
  Ben-Gurion University of the Negev,
  Beer-Sheva 84105, Israel \\
  $^4$The Ilse Katz Center for Nano-Science,
  Ben-Gurion University of the Negev,
  Beer-Sheva 84105, Israel
}

\begin{abstract}
We briefly review the theoretical formulations and applications of the Aharonov--Bohm effect and the Aharonov--Casher effect with emphasis on mesoscopic physics. 
Topics relating to the Aharonov--Bohm effect include: locality, periodicity, non-integrable phase factors, Abelian gauge theory, interference, the spectrum and persistent current of electrons on a ring pierced by a magnetic field, Onsager reciprocity relations, and Aharonov--Bohm interferometer.
Topics relating to the Aharonov--Casher effect include: a magnetic dipole in an electric field, locality, periodicity, non-Abelian gauge invariance, SU(2) non-integrable phase factors, spin-orbit coupling, Pauli equation, Rashba Hamiltonian, Aharonov--Casher interferometer, conductance and polarization in two-channel systems due to the Aharonov--Casher effect.

%

\end{abstract}
\maketitle



 \section{Introduction}
The Aharonov--Bohm effect (Aharonov Bohm 1959) and Aharonov--Casher effect (Aharonov Casher 1984) are purely quantum mechanical in nature.  They highlight the crucial role of the phase of the wave function and its experimental significance (Berry 1984, Loss 1990, Stern {\it{et al}}. 1990, Stern, 1992, Aronov 1993, Qian 1994).
The Aharonov--Bohm effect shows that a charged particle is affected by the magnetic field even if the particle is confined to a region in which  the magnetic field ${\bf B}$ is absent.  Thus, the magnetic field acts non-locally on charged particles, and the fundamental pertinent field affecting charged particle is the vector potential ${\bf A}$, and the relevant theory is characterized by Abelian gauge invariance. 
The Aharonov--Casher effect applies to particles with an intrinsic magnetic moment (they need not necessarily be charged), whose (spinor) wave function acquires a matrix valued phase factor. Consequently, the relevant theory is characterized by non-Abelian gauge invariance (Yang and Mills 1954). Several examples relevant to condensed matter physics are discussed, mainly for mesoscopic systems (Stone 1985, Lee and Stone 1985, Imry 1997, Akkermans and Montambaux 2007), supported by short derivations of the basic equations that model the observed phenomena.\\

\section{Aharonov--Bohm effect}  \label{AB}

The Aharonov--Bohm effect (Aharonov and Bohm 1959, Aharonov and Bohm 1961, Aharonov and Anandan 1987), is a spectacular manifestation of the difference between classical and quantum theories of charged particles interacting with electromagnetic forces; it stresses the crucial role of the phase of the charged particle's wave function.  It predicts that the behavior of a charged particle is affected by the electromagnetic field even when the particle is always in a region where the field vanishes.  Shortly after being predicted, it was first confirmed experimentally (Chambers 1960). An experimental verification in a toroidal magnetic field was reported in (Osakabe {\it{et al}} 1986) and its verification in carbon nano-tubes was reported in (Bachtold {\it{et al}}. 1999). It was also confirmed in tunneling experiments, (Atsushi {\it{et al}}. 2014) and tomography  (Valagiannopoulos et al. 2018). 

In classical electrodynamics, the electric and magnetic fields ${\bf E}$ and ${\bf B}$ fully describe the electromagnetic field and the phenomena involving the interaction of matter and photons. The vector potential ${\bf A}$  is introduced as an auxiliary field in order to solve Maxwell's equations.   In quantum mechanics, the Aharonov--Bohm effect shows that this is not the case. For charged particles, knowing ${\bf B}$ (assume ${\bf E}=0$ for simplicity), does not completely determine all the electromagnetic effects on the particle. Experimental and theoretical consequences of the Aharonov--Bohm effect were elaborated in numerous papers and books, see e.g., (Olariu and Popescu 1985, Peshkin and Tonomura 1989, Tonomura 2006).

 In order to describe the effect in simple terms, let us consider, as in Fig.~\ref{Fig1}, a two slit interference pattern of electrons moving in the $x$-$y$ plane.  The plane is pierced by a circular tube of radius $R$ centered at the origin, namely $\sqrt{x^2+y^2}<R$, in which there is a uniform perpendicular magnetic field, ${\bf B}=B_z \Theta(R-r)$ (see the small circle with $B$ in Fig.~\ref{Fig1}). Outside the tube ${\bf B}=0$ yet ${\bf A} \ne 0$ (but nevertheless, the magnetic field $\nabla {\times} {\bf A}= 0$), see (Tonomura 2006, Batelaan and Tonomura 2009). The electron wave function is $\psi({\bf r})$ and the density pattern   on the screen on the right is proportional to $\vert \psi({\bf r})\vert^2$. The line integral $\oint {\bf A}  \cdot d {\bf r} =\pi R^2 B \ne 0$ along {\it{any closed curve encircling the tube}}, equals the magnetic flux through the tube.   Because of the Aharonov--Bohm effect, experiments performed on electrons that are located outside the flux tube (where ${\bf B}=0$), e.g., measuring the density on the screen), yield results that depend solely on the (dimensionless) quantity, 
\begin{equation} \label{ABphase}
\Phi_{\text AB} \equiv \frac{e}{\hbar c} \oint {\bf A}  \cdot d {\bf r}  \ne 0.
\end{equation}
 This quantity is called the Aharonov-Bohm phase.  In a sense, the magnetic field acts where it is not present (see however Avishai {\it{et al}}. 1972, Roy 1980, Vaidman 2008, Popescu 2010, Vaidman 2012, Pearle and Rizzi 2017, Heras 2022). In other words, knowledge of the scalar potential $V$ and the vector potential {\bf $A$} contain more information than the information about the fields ${\bf E}$ and ${\bf B}$. The Aharonov--Bohm effect can also test the role of the gravitational potential versus that of the gravitational field (Dowker 1967,  Hohensee {\it{et al}}. 2012, Overstreet {\it{et al}}. 2022).

\subsection{Periodicity in the Aharonov-Bohm phase}
As predicted by Aharonov and Bohm and substantiated shortly after their publication (Byers and Yang 1961), such experimental results are periodic functions of $\Phi_{\text AB}$ with period
\begin{equation} \label{Phi0} 
\Phi_0 \equiv \frac{hc}{e}
=4.1414 \times 10^{-7} \, \mbox{Gauss-cm}^2 ,
\end{equation}
which serves as a  quantum unit of flux.  
Consequently, (Wu and Yang 1975), introduced the concept of {\it{phase factor}}, 
\begin{equation} \label{ABPF}
\Lambda_{\text AB} \equiv e^{i \Phi_{\text AB}},
\end{equation}
and stressed that, although $\Phi_{\text AB}$ contains more information than $\Lambda_{\text AB}$, this extra information is not measurable. In other words, the result of experiments depend solely on the  phase factor $\Lambda_{\text AB}$.  If there are two cases $a$ and $b$ such that $\Phi_{\text AB}^a - \Phi_{\text AB}^b=n\frac{hc}{e}$ ($n$ = integer), no experiment performed on an electron {\it{outside the tube}} can distinguish between the two cases.

\subsection{Gauge invariance}
The proof of the statement above (Wu and Yang 1975) proceeds by applying a (unitary) {\it{gauge transformation}} between the corresponding electron wave functions, $\psi_b({\bf r})=e^{i \alpha({\bf r})}\psi_a({\bf r}) \equiv S({\bf r})\psi_a({\bf r})$, implying ${\bf A}_b ({\bf r})-{\bf A}_a({\bf r})=\frac{\hbar c}{e} {\bf \nabla}\alpha({\bf r})$. Since the curl of the left hand side vanishes, the solution for $\alpha({\bf r})$ is, $ \alpha({\bf r})=\frac{e}{\hbar c} \int_{{\bf r}_0}^{\bf r}[{\bf A}_b ({\bf r})-{\bf A}_a({\bf r})]\cdot d {\bf r}$, where ${\bf r}$ and ${\bf r}_0$ are two points along the chosen curve outside the tube.  Therefore, 
\begin{equation} \label{alpha}
\frac{e}{\hbar c} \oint [{\bf A}_b ({\bf r})-{\bf A}_a({\bf r})]\cdot d {\bf r}=\Phi_{\text AB}^b-\Phi_{\text AB}^a=2n \pi ,
\end{equation}
hence, $[\Lambda_{\text AB}]_a=[\Lambda_{\text AB}]_b$.  In the language of (Wu and Yang 1975), {\it{the two cases are  gauge transformable into each other and no physically observable effects can differentiate between them}}. 

The concept of gauge transformation is generalized  to the case where the line integral is performed for any curve between ${\bf r}_0$ and ${\bf r}$ by introducing a {\it{non-integrable (path dependent) phase factor}},
\begin{equation} \label{nipf}
  \Lambda_{pq}\equiv e^{\frac{ie}{\hbar c} \int_{{\bf r}_p}^{{\bf r}_q}{\bf A}({\bf r})\cdot d {\bf r}},
\end{equation}
such that an arbitrary gauge transformation $e^{\frac{ie}{\hbar c} f({\bf r}_p)} \Lambda_{pq}e^{\frac{-ie}{\hbar c} f({\bf r}_q)}$, does not affect the result of measurements.  Here $f(\bullet)$ is  any continuous function taking ${\bf r}_p$ outside the tube to ${\bf r}_q$ outside the  tube, and the factor $e^{\frac{ie}{\hbar c} f({\bf r})}$ is a transformation of the wave-function $\psi({\bf r})$. Wu and Yang then conclude that {\it{electromagnetism is the gauge-invariant manifestation of non-integrable phase factors}}.

\subsection{Examples in condensed matter physics}
In the Aharonov--Bohm effect phase coherence is a crucial factor. Hence, pertinent experiments must be carried out at low temperature and with systems of small size (up to a few microns). The underlying physics of systems whose size is larger than the atomic size  but much smaller than macroscopic size   is termed {\it{mesoscopic physics}} (Lee and Stone 1985, Stone 1985, Imry 1997). In the examples discussed below, the spin of the electron does not play a role. The Aharonov--Bohm effect for particles with spin is discussed in (Hagen 1990a,  Nitta 1997,  Morpurgo 1998, Nitta 1999, Yau 2002, Grbic 2007).

\subsubsection{Interference}
\begin{figure}[htb]
\centering
{\includegraphics[width=0.7\linewidth]{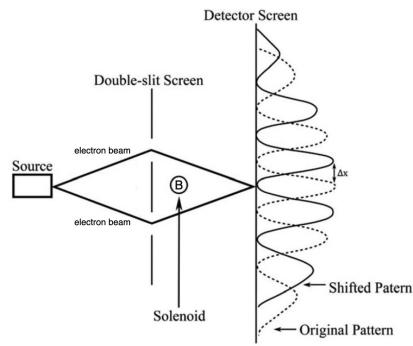}} 
\caption{\footnotesize 
Aharonov--Bohm interference experiment (schematic): A coherent electron beam emanating for the source propagates in a region where there is no magnetic field but is affected by a flux tube of radius $R$ (with the electron beam not able to enter the tube) within which the magnetic field ${\bf B}$ (along the axis of the tube) is finite and uniform, hence the magnetic flux is $\Phi_{\text AB}=\pi R^2 B$. The interference pattern on the screen is a periodic function of $\Phi_{\text AB}$ with period $\frac{hc}{e}$.  The original pattern (bold line) is obtained for $\Phi_{\text AB}  = n \Phi_0$ and the shifted pattern (dashed line) is obtained for $\Phi_{\text AB} \ne n \Phi_0$ ($n=$ integer). (Modified from Shech 2022, Fig.~2). 
}
\label{Fig1}
\end{figure}

Figure \ref{Fig1} is a schematic description of an Aharonov--Bohm interference wherein 
a beam of electrons propagates toward a screen. The electrons avoid the tube through which a magnetic field generates a magnetic flux $\Phi_{\text AB}$. Outside the tube the magnetic field vanishes but the vector potential ${\bf A}$ is finite. The electrons do not feel the magnetic field, but on going outside the tube from a point $q$ to a point $p$ the wave function of the electron gains a phase as given in Eq.~(\ref{nipf}). This leads to an interference pattern whose intensity is shown on the screen. The intensity pattern is periodic in $\Phi_{\text AB}$ with period $\Phi_0=\frac{hc}{e}$.  
 
\subsubsection{Electrons on a ring}
Impressive progress in experimental techniques have enabled fabrication of micron-sized conducting rings (Webb {\it{et al}}. 1985, Imry and Webb 1986).  Specifically, consider uniformly distributed electrons propagating in a small metallic ring of radius $R$ and length $L=2 \pi R$. The ring is placed on a plane and pierced by a perpendicular magnetic field ${\bf B}$ such that the magnetic flux is $\Phi_{\text AB}=\pi R^2 B$, and the corresponding vector potential ${\bf A}=\frac{\Phi_{\text AB}}{L}\hat{\bm \theta}$ is along the ring (the magnetic field acting on the circumference of the ring vanishes). Here the electrons are treated as non-interacting spin-less fermions of charge $-e$ and mass $m$ (see however, Hagen 1990a).  Due to the Aharonov--Bohm effect, this simple system exhibits beautiful phenomena both in equilibrium and, if coupled to leads with a small potential difference, out of equilibrium (see the section Aharonov--Bohm interferometer below).

\subsubsection{Spectrum}
\begin{figure}[!ht]
\centering
\includegraphics[width=0.45\textwidth]{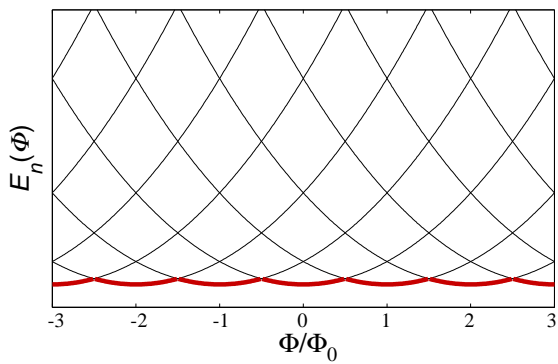}
\caption{Energy eigenvalues $E_n(\Phi_{\text AB})$ versus magnetic flux
$\Phi_{\text AB}/\Phi_0$, taken from (Band and Avishai 2012).  Level crossings occur at $\Phi_{\text AB} = m \Phi_0$ and $\Phi_{\text AB} = (m+1/2) \Phi_0$ in a clean system ($m$=integer).  However, any amount of disorder or imperfection leads to avoided crossings and a pattern of energy bands with small gaps between them.  For any fixed value of the flux, the levels should be ordered from low energy to higher energy.  The ground-state energy of a single electron is shown as the thick red line.  With this ordering, the spectrum is a periodic function of the flux
with period $\Phi_0$.}
\label{Fig2}
\end{figure}

Let $0 \le x \le L$ denote an electron coordinate along the ring. The time-independent Schr\"{o}dinger equation for electrons (charge $-e$ and mass $m_e$) moving on the ring is,
\begin{equation}  \label{Hamiltonian}
H \psi = \frac{1}{2m_e}  \left (p +\frac {e}{c}\frac{\Phi_{\text AB}} {L} \right)^2 \psi= \veps \psi ~,
\end{equation}
where $p = - i \hbar \frac {d}{d x}$ is the momentum operator and $c$ is the speed of light.  Note that $[p,H]=0$ since $\Phi_{\text AB}$ is independent of $x$.  The periodic solutions are plane waves, $\psi(x) = L^{-1/2} \, e^{ikx}$.  Since the wave function must be single valued, the wave-numbers are quantized, $k=k_n=2 n \pi/L=n/R$, with $n=0,\pm1,\pm 2 \ldots$.  Substituting this wave function into Eq.~(\ref{Hamiltonian}), one obtains the energy eigenvalues $E_n$,
\begin{eqnarray}  \label{ePhi}
     \veps_n(\Phi_{\text AB}) &=& \frac{1}{2m_e} \left (\hbar k_n + \frac{e\Phi_{\text AB}} {cL}
     \right )^2 \\ \nonumber
     &=& \frac{\hbar^2} {2m_eR^2} \left(n + \frac{\Phi_{\text AB}}{\Phi_0} \right )^2 ~,
\end{eqnarray}
Note that $\veps_n(\Phi_{\text AB})$ is not periodic in  $\Phi_{\text AB}$ and also, $\veps_n(\Phi_{\text AB}) \ne \veps_n(-\Phi_{\text AB})$ (which is physically wrong). Using the degeneracy $\veps_{n+1}(\Phi_{\text AB})=\veps_{n}(\Phi_{\text AB}+\Phi_0)$ we can  reorder the quantum numbers $\{n \in \mathbb{Z} \}$ and the energies $\{ \veps_n(\Phi_{\text AB}) \}$ such that (see Fig.~\ref{Fig2}), 
\begin{eqnarray} \label{reorder}
&&0 \le E_n(\Phi_{\text AB}) \le E_{n+1}(\Phi_{\text AB}), \\ 
&&E_n(\Phi_{\text AB})=E_n(\Phi_{\text AB} \pm m \Phi_0)=E_n(-\Phi_{\text AB}),  \\
&& {\mbox{where }} n,m=0,1,2 \ldots . \nonumber
\end{eqnarray}  

At zero temperature, the ground-state energy is given by the sum of single-particle energies of levels below the Fermi energy $E_F$,
\begin{equation} \label{GSPHI}
  E_{\mathrm{gs}}(\Phi_{\text AB}) = 2 \sum_{E_n(\Phi_{\text AB}) < E_F} E_n(\Phi_{\text AB}).
\end{equation}
where the factor 2 is due to spin degeneracy. 
At higher temperature, the levels are occupied according to the Fermi-Dirac distribution function.

\subsubsection{Persistent Current}
One of the important consequences of the Aharonov--Bohm effect in the
ring geometry is that there can exist a non-zero steady-state current
in the ring. This was first suggested in (B\"uttiker {\it et al}. 1983),
and was observed experimentally several years later (Levy {\it et al}. 1990). 
The existence of the current is related to the fact that in a magnetic field,
time-reversal symmetry is violated and the Hamiltonian is complex (albeit Hermitian).
This current is not the usual conduction current where electrons in a
metal at the Fermi energy exchange energy with an external electric
field.  Rather, it is an equilibrium property
of the system, and there is no dissipation associated with this current.
For the moment, confining our discussion to zero temperature, all the electrons in the levels below the Fermi energy $E_{\text F}$ contribute to the current.  The current operator can be written as 
\begin{equation}  \label{Eq:MSI.AB.4}
     \hat{I} =  -c \frac{\partial H}{\partial \Phi_{\text AB}} = \frac{-e}{L}
     \left[\frac{1}{m}(p+\frac{e\Phi_{\text AB}}{cL})\right] = \frac{-e}{L}
     \hat{v} ~,
\end{equation}
where ${\hat v}$ is the velocity operator.  The contribution of level $n$ below the Fermi energy to the current is obtained by taking the expectation value of ${\hat I}$ with the wave function $\psi_n$.  Using the Feynman-Hellman theorem, we can to equate the expectation  $\langle \psi_n \vert \frac{\partial H}{\partial \Phi_{\text AB}}\vert \psi_n \rangle$ to $\frac{\partial E_n(\Phi_{\text AB})}{\partial \Phi_{\text AB}}$. The total current at zero temperature is then given by,
\begin{equation}  \label{Total-current}
    I(\Phi_{\text AB}) =  -c \sum_{E_n(\Phi_{\text AB}) < E_F} \frac{\partial E_n(\Phi_{\text AB})}{\partial \Phi_{\text AB}} = -c \frac{d E_{\mathrm{gs}}}{d \Phi_{\text AB}} ~.
\end{equation}
As a result of the properties of the spectrum listed above, it is
clear that the current is a periodic function of $\Phi_{\text AB}$ with period
$\Phi_0$ and is an antisymmetric function of the flux,
\begin{equation} \label{I-periodic}
    I (\Phi_{\text AB}) = -I(-\Phi_{\text AB}) ~, \quad  I(\Phi_{\text AB}+m \Phi_0) = I(\Phi_{\text AB}) ~.
\end{equation}

The above analysis is pertinent to a one dimensional metallic ring geometry. It is worthwhile  mentioning other geometries.
Persistent current is analyzed  for a Corbino disk geometry (Avishai et al. 1993,  Avishai. and Kohmoto 1993, Yerin et. al. 2021), and for a closed flux line of arbitrary shape and size (Heras 2022).  The Aharonov--Bohm effect for an electron on a spherical shell was discussed be (Wu and Yang 1975), 
 
What happens if the ring is not ``clean'', e.g., if a disordered potential is present in the ring? (Altshuler 1981, Imry 1997).  In this case, the electrons can be scattered elastically from numerous fixed scattering centers.  Such scattering centers are related to imperfections and impurities.  Experimentally, they are often unavoidable.  In many cases, the positions and potential strengths of these impurities are generically random; this kind of potential is called {\it quenched disorder}.  The main consequences of
the presence of quenched disorder in the ring are as follows:
\begin{enumerate}
\item{} Level crossings shown in Fig.~\ref{Fig2} are
avoided and the dependence of energy levels on the flux is smooth.
\item{} The periodicity and symmetry relations of the energy and
current remain unchanged.  The Fourier expansion of the current is,
\begin{equation} \label{Eq:MSI.AB.3e}
    I = \sum_{k=1}^{\infty} I_{k} \sin(2 \pi k \Phi_{\text AB}/\Phi_0) ~.
\end{equation}
In some experiments, the current is found to have an effective period
of $\Phi_0/2$ because, in the presence of weak disorder,
 the dominant Fourier components have even $k$ (Imry 1997). 
\item{} 
In the presence of quenched disorder, the wave
functions in one dimension decay exponentially at long distance with
some characteristic length scale $\xi$ which depends on energy and
strength of disorder, and referred to as the {\it localization
length}.  This phenomena is termed as {\it Anderson localization} (Anderson 1958).  
Thus, if the ring size is such that $L>\xi$, we expect the current to decay exponentially with $L$.
\item{} For $L < \xi$, the current persists and is
measurable.  
\item{} The order of magnitude of persistent current in the presence of quenched disorder depends on the experimental situation. There are experiments on a single ring and then one speaks of a {\it sample-specific} persistent current.  In this case the order of magnitude is $E_c/\Phi_0$, where according to the Thouless relation, $E_c = \hbar D/\ell^2$ (Eduards and Thouless 1972; Altland {\it{et al}}. 1996), where $D$ is the diffusion constant and $\ell$ is a typical random walk step.

On the other hand, experiments involving hundreds of rings are referred to as {\it impurity-ensemble} experiments. The contribution of many currents tends to cancel out, since it has, for example, a random slope at the origin ($\Phi_{\text AB} = 0$). Thus, ensemble averaging is expected to reduce the magnitude of the persistent currents. Calculations pertaining to weak disorder performed at finite temperature within the grand canonical ensemble (Entin-Wohlman and Gefen 1989), conclude that after sample averaging, the persistent current virtually vanishes. ``Grand canonical'' means in this connection that the rings are taken to be connected to a particle bath at temperature $T$, having a chemical potential $\mu$, and the electron number $N$ may change with $\Phi_{\text AB}$ in each ring.
\end{enumerate}

An important aspect of the behavior of such systems is how the interaction between electrons affects the persistent current.  Numerical calculations (Berkovits and Avishai 1995, Berkovits and Avishai 1996) reveal significant interaction-induced enhancement of persistent currents in 2D disordered cylinders.  In the paper by (Entin-Wohlman {\it{et al}}. 2003) it is shown that persistent currents in interacting Aharonov--Bohm interferometers
are enhanced by acoustic radiation. See also (Zvyagin 2020).

\subsubsection{Aharonov--Bohm interferometer}
Electron-based interferometry is a powerful device for probing  coherent quantum phenomena in condensed matter physics.  An example of the relevance of the Aharonov--Bohm effect in a  
non-equilibrium situation is an electron interferometer, which is experimentally 
easier to control than an isolated ring.  
The measured quantities are conductance and current, that are easier to measure than the persistent current.
 The underlying 
tool used to calculate these observables is quantum scattering theory. The Aharonov--Bohm interferometer highlights 
some fundamental concepts in quantum mechanics such as interference,
coherence, gauge invariance, current conservation and 
Onsager reciprocity relations (Onsager 1931).  

\begin{figure}[!ht]
\centering
\includegraphics[angle=0,width=0.45\textwidth]{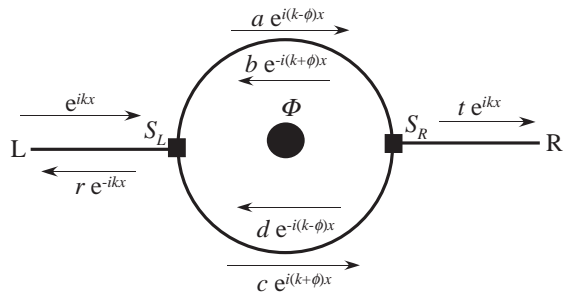}
\caption{Schematic illustration of scattering through an Aharonov--Bohm
interferometer. Modified from Imry 1997, page 110 Fig.  5.4. }
\label{Fig3}
\end{figure}

In its simplest version, an Aharonov--Bohm interferometer (see
Fig.~\ref{Fig3}) consists of two one dimensional ideal conductors $L$ and $R$
(for left and right) connected by three port contacts $S_L$ and $S_R$,
sometimes referred to as {\it splitters}, to an ideal ring-shaped
conductor of radius $\rho$ threaded by a central magnetic flux.
The conductors are connected on their other side to electron reservoirs at potentials 
$V_L \gtrapprox V_R$ so there is a small potential difference between the left and right reservoirs, 
and a small current from left to right. 
There is a single incoming channel and a single outgoing channel, and
therefore, according to the Landauer formula (Landauer 1957, B\"uttiker {\it{et al}}. 1985), 
the conductance of the interferometer (in units
of the quantum conductance $G_0 \equiv e^2/h$), is equal to the transmission coefficient.  
%
The magnetic field vanishes everywhere except at the small aperture marked as black circle.
The electrons stay on the wires of the interferometer and do not reach the region
where the magnetic field is present.  Nevertheless, the transmission
(and hence the conductance) depends on the magnetic flux.  This is a
purely quantum mechanical effect with no classical analog.
%

To describe the underlying physics it is useful to introduce dimensionless variables using the radius
$\rho$ of the ring as the unit of length.  The dimensionless
coordinate is $x$ (in units of $\rho$), the wave number is $k$ (in
units of $1/\rho$), the dimensionless energy is $\varepsilon =
\frac {2 m E \rho^{2}} {\hbar^{2}} = k^{2}$, (where $E$ is the energy in physical units), and the dimensionless magnetic flux is $\phi = \Phi_{\text AB}/\Phi_{0}$, where $\Phi_{0} \equiv \frac{hc}{e}$. The 
incoming plane wave with (dimensionless) energy $\veps = k^{2}$ propagates
toward the interferometer from the left and is partially reflected at $S_L$ leftward with
reflection amplitude $r$ and partially transmitted into the two arms of the ring.  After traveling through the two arms of the ring it is partially transmitted at $S_R$ with
transmission amplitude $t$ at partially reflected back to the arms of the ring.  The goal is to calculate the transmission and hence, according  to the Landauer formula (Landauer 1957), the conductance 
$G(k,\Phi_{\text AB}) = 2\frac {e^{2}}{h}|t|^2$.  The external flux is a convenient control
parameter, and the dependence of $G(k,\Phi_{\text AB})$ on energy and magnetic
flux reflects the roles of quantum interference (at the splitters) and gauge
invariance (the periodicity of $G(k,\Phi_{\text AB})$).  Note that the system cannot be regarded as purely
one dimensional since part of it (the ring with the flux) is not
simply connected.  The calculation requires finding the two linearly
independent solutions on every 1D arm of the ring and matching
the solutions at the two splitters.  The motion of the electron on the
ring is governed by the Schr\"odinger equation in the presence of
magnetic flux.
The variable $x$ in Fig.~\ref{Fig3} assumes values $-\infty < x
\le 0$ on $L$, $0 \le x < \infty$ on $R$ and $0 \le x \le \pi$ on
either arm of the ring where $x = 0$ at $S_L$.  Gauge invariance
requires that all measurable quantities are periodic functions of
$\phi$ with period $1$, and Onsager reciprocity relations require that
the conductance is symmetric with respect to $\phi$ (see
Eq.~\ref{Eq:Dot.ByersYang}). 
Thus, we have,
\begin{subequations}  \label{Eq:Dot.SEring}
\begin{equation}
    -\frac{d^{2}}{dx^{2}} \psi(x) = k^{2} \psi(x), \ \ \ (x < S_L, \ \
    \ x>S_R) ~,
\end{equation}
\begin{equation}
    (-i \frac{d}{dx}+\phi)^{2} \psi(x) = k^{2} \psi(x), \ \ \ ( S_L
    \le x \le S_R) ~.
\end{equation}
\end{subequations}
The solutions corresponding to an incoming wave from the left (see Fig.~\ref{Fig3}) are,
\begin{subequations}
\begin{equation}
    \psi(x) = e^{ikx}+r \, e^{-ikx} \ \ \ (x<S_L) ~,
\end{equation}
\begin{equation}
    \psi(x) = t \, e^{-ikx} \ \ \ \ \ (x>S_R) ~,
\end{equation}
\begin{eqnarray}
    \psi(x) = a \, e^{i (k-\phi)x} + b \, e^{-i (k+\phi)x} \\ \nonumber
    \ (S_L \le x \le S_R, \ {\mathrm{upper \, arm}}),
\end{eqnarray}
\begin{eqnarray}
    \psi(x) = c \, e^{i (k+\phi)x} + d \, e^{-i (k-\phi)x} \\ \nonumber
    \ (S_L \le x \le S_R, \ {\mathrm{lower \, arm}}).
\end{eqnarray}
\end{subequations}
The junctions at points $S_L$ and $S_R$ are referred to as splitters.
The precise relations between the incoming and outgoing amplitudes at
each splitter depend on the detailed structure of the device.  The
only restriction is that current should be conserved at each splitter.
To satisfy this restriction, we characterize the splitters $S_L$ and
$S_R$ by two unitary 3$\times$3 $S$ matrices ${\cal S}_L$ and ${\cal
S}_R$ relating the three amplitudes of the incoming waves in the
junction to the three amplitudes of the outgoing waves at this
junction.  The form of these $S$ matrices is dictated by the requirements of 
wave function continuity and current conservation but otherwise they are dependent on the
structure of the contacts and the experimental setup.  Continuity and current conservation
at the two junctions yields,
\begin{eqnarray}  \label{Eq:Dot.matchingAB}
	&&{\cal S}_L \begin{pmatrix} 1 \\ b \\ d \end{pmatrix} =
	\begin{pmatrix} r \\ a \\ c \end{pmatrix}, \\ \nonumber
	&&{\cal S}_R \begin{pmatrix} 0 \\ ae^{i (k-\phi) \pi} \\ c e^{i
	(k+\phi) \pi}\end{pmatrix} =\begin{pmatrix} t \\ be^{-i
	(k+\phi) \pi} \\ d e^{-i (k-\phi) \pi} \end{pmatrix}.
\end{eqnarray}
These are six linear inhomogeneous equations for the six unknowns
$a,b,c,d,r$, and $t$, which can be solved once the unitary matrices
${\cal S}_L$ and ${\cal S}_R$ are given.  The solution must satisfy 
continuity of the wave function at the junction and obey the
current conservation constraint (Kirchoff's law). These constraints 
determine the form of the unitary matrices in the splitters,  (Shapiro 1983),
\begin{equation} \label{Smattrix}
{\cal S}_L={\cal S}_R= \begin{pmatrix} 0&-\tfrac{1}{\sqrt{2}}&-\tfrac{1}{\sqrt{2}}\\-\tfrac{1}{\sqrt{2}}&\tfrac{1}{2}&-\tfrac{1}{2}\\
-\tfrac{1}{\sqrt{2}}&-\tfrac{1}{2}&\tfrac{1}{2} \end{pmatrix}
\end{equation}
 Current conservation implies that $\vert t \vert^2+\vert r \vert^2=1$. Another test is
that the conductance through the interferometer $G(\phi) = 2
\frac{e^{2}}{h} |t(\phi)|^{2}$ (the factor 2 is due to spin degeneracy), should be a symmetric and periodic
function of $\phi$ with period 1,
\begin{equation}  \label{Eq:Dot.ByersYang}
    G(\phi) = G(-\phi) ~, \quad    G(\phi) = G(\phi+1) ~.
\end{equation}
The first equality is a special case of a more general set
of constraints resulting from the combination of time-reversal
invariance and magnetic field reversal, known as {\it Onsager
relations} (Onsager 1931).  The second relation,
the periodicity of $G(\phi)$, is a consequence of gauge invariance,
as demonstrated  in the {\it Byers-Yang theorem} (Byers and Yang 1961, Bloch 1970).   
As already discussed above, it states that every physical quantity that is measured in a system
which is not simply connected and threaded by a magnetic flux $\Phi_{\text AB}$
is a periodic function of $\Phi_{\text AB}$ with period given by the flux quantum
$\Phi_{0} = h c/e$. 


This system is a beautiful manifestation of the  Aharonov--Bohm effect, and, in addition, reveals numerous fundamental concepts in quantum mechanics (Gefen {\it{et al}}. 1984, van Oudenaarden {\it{et al}}. 1998, Cheung {\it{et al}}. 1989, Entin-Wohlman and Gefen 1989, Aharony {\it{et al}}. 2002, Aharony {\it{et al}}. 2003, Entin-Wohlman {\it{et al}}. 2004, Aharony {\it{et al}}. 2005, Aharony {\it{et al}}. 2006, Aharony {\it{et al}}. 2014, Aharony {\it{et al}}. 2019). The example analized above involves only two-port systems. The scattering theory and Aharonov--Bohm effect in multi-port systems has been developed in (B\"uttiker  1985, B\"uttiker  1986,    
Imry 1986, Avishai and Band 1987, B\"uttiker  1988,  B\"uttiker  1990,  B\"uttiker  1992, Cernicchiaro 1997). 

The Aharonov--Bohm effect was predicted to be a possible tool for identifying the set of sites visited by a wave packet, such that for particular values of the magnetic flux, the wave packet is bound, and this effect is referred to as Aharonov--Bohm caging (Vidal {\it et al}. 1998).  This effect was recently observed in cold atom systems (Li {\it et al}. 2022).  Peculiar aspects of the Aharonov--Bohm effect in graphene are reported by (Recher {\it et al}. 2007, Rycerz and Beenakker 2007), and in molecular physics by (Sjoqvist 2014). Recently, two experiments were performed manifesting the Aharonov--Bohm effect in the study of anyons, (Ronen {\it{et al}}, 2021) using a Fabry--P\'erot quantum Hall interferometer, 
and (Hemanta {\it et al} 2023) using a Mach-Zehnder interferometer.  Both experiments exposed the unconventional exchange statistics of these exotic quasiparticles which occur only in two-dimensional systems, and play a crucial role in the fractional quantum Hall effect.  The latter experiment also studies the braiding phases of anyons.

\section{Aharonov--Casher Effect}  \label{AC}
\begin{figure}[!ht]
\centering
\includegraphics[width=0.45\textwidth]{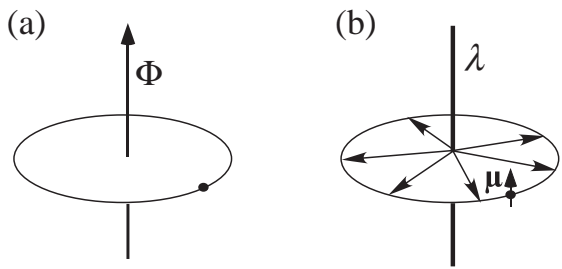}
\caption{A schematic illustration of the Aharonov--Bohm and Aharonov--Casher
effects.  (a)  Aharonov--Bohm Effect: An electron moving on a metallic ring threaded by a
magnetic flux $\Phi_{\text AB}$.  When the electron encircles the ring, its wave
function gains a phase $e^{2 \pi i \phi }$, where $\phi=\Phi_{\text AB}/\Phi_0$.
The spin of the electron does not play a role.  (b)  Aharonov--Casher Effect: An electron with
magnetic moment ${\boldsymbol \mu}$ moving on a metallic ring threaded
by a charged wire of longitudinal charge density $\lambda$.  This
results in a radial electric field ${\bf E} = \frac{\lambda}{L}{\hat
{\bf r}}$ (black arrows), where $R$ is the radius of the ring.  When
the electron encircles the ring, its wave function (a two component
spinor) is multiplied by a unitary matrix $e^{2 \pi i \gamma \sigma_z}$ 
where $\gamma=\frac{g \mu_B \lambda}{2 h c}$.  The charge of the electron
does not play a role here. Taken from (Band and Avishai 2012).} 
\label{Fig4}
\end{figure}

The Aharonov--Casher effect (Aharonov and Casher 1984, Aharonov {\it{et al}}. 1988, Bergsten 2006, Rohrlich 2009) is an astounding quantum phenomena  which is related to the Aharonov--Bohm effect. 
The relation is sketched in Fig.~\ref{Fig4}. The Aharonov--Casher effect occurs when a particle with non-zero magnetic moment ${\boldsymbol \mu} = g \mu_B {\bf S}/\hbar$ moves in a doubly connected region and is affected by an {\it{electric field}} that leads to spin-orbit interaction. The particle need not be charged.  It can be an electron (K\"oenig {\it{et al}}. 2006), a neutron (Cimmino {\it{et al}}. 1989),  a magnetic vortex (Elion {\it{et al}}. 1993) or a neutral atom with non-zero spin.  An essential point is that no force acts on the particle. In condensed matter physics  the effect is mainly relevant for electrons (Mathur and Stone  1992, Bogachek and Landman 1994, Frustaglia 2004, Saarikosk 2015, Frustaglia 2020). 

First we consider a simple case wherein an electron moves on a ring threaded through its center by a charged line of charge density $\lambda$ which is perpendicular to the plane of the ring.   The setup is shown in Fig.~\ref{Fig4}(b), where a comparison of the ring geometries of the Aharonov--Bohm [see Fig. \ref{Fig4}(a)] and Aharonov--Casher effects is illustrated.

A heuristic explanation of the role of spin-orbit in the  Aharonov--Casher effect (Aharonov and Casher 1984) for charged particles, e.g. electrons, goes as follows:  A particle (say, an electron of mass $m_e$), subject to an electric field ${\bf E}$ feels, in its rest frame, an effective magnetic field ${\bf H}_{\mathrm{eff}} = ({\bf v}/c) \times {\bf E} = \frac{1}{m_ec} {\bf p} \times {\bf E}$, where ${\bf v}$ is the particle's velocity and ${\bf p}$ is its momentum.  This results in an effective Zeeman Hamiltonian, $H_Z \equiv -\ {\boldsymbol \mu} \cdot {\bf H}_{\mathrm{eff}} = g (\mu_B/m_ec) ({\bf S}/\hbar) \cdot {\bf p} \times {\bf E}$, where $g \approx 2$ reflects the anomalous magnetic moment of the electron.  For an electron subject to an electric field ${\bf E}$ the spin-orbit term of the Hamiltonian is written (in proper Hermitian form) as,
\begin{equation} \label{Eq:MSI.Hso}
    H_{\mathrm{so}} = \eta ({\bf p} \cdot {\bm \sigma} \times
    {\bf E} + { \bm \sigma} \times {\bf E} \cdot {\bf p} ) ~,
\end{equation}
where $\eta=\frac{g \mu_B}{4 m_ec}$.  For electrons, 
the Hamiltonian $H_{\mathrm{so}}$, appears in the  Pauli
Hamiltonian (Pauli 1927) which arises in the context of working out the
non-relativistic limit of the Dirac equation by expanding it in the
Foldy-Wouthuysen scheme (Foldy 1950).  It  is responsible for a myriad of 
phenomena. In condensed matter physics this Hamiltonian is 
related to the Rashba Hamiltonian (Rashba 2006, Engel {\it{et al}}. 2006). 
In contrast with the Hamiltonian $H$ defined in Eq.~(\ref{Hamiltonian}), 
the Hamiltonian $H_{\mathrm{so}}$ is invariant under time-reversal that 
reverses the signs of both ${\bfp}$ and ${\bm \sigma}$.

A simple example of the Aharonov--Casher effect is illustrated by considering the ring
geometry shown in Fig.~\ref{Fig4}(b).  An electron
moving on a ring of radius $R$ experiences a static electric field
${\bf E}=\frac{\lambda}{L}{\hat {\bf r}}$ produced by a uniformly charged
wire with constant charge per unit length $\lambda$, stretched along
the $z$ axis passing through the ring's center (here, $L = 2\pi R$).
It is instructive to write the Hamiltonian in a form similar to that of Eq.~(\ref{Hamiltonian}),
\begin{equation} \label{ACRing}
    {\cal H} = \frac{{\bf p}^2}{2m_e}+H_{\mathrm{so}} =
    \frac{1}{2m_e}({\bf p}+\frac{g \mu_B}{2 c} {\bm \sigma} \times {\bf
    E})^2 - \frac{(g \mu_B {\bf E})^2}{8 m_e c^2} ~.
\end{equation}
The term $({\bf p}+\frac{g \mu_B}{2 c} {\bm \sigma} \times {\bf E})$ is the {\it{covariant SU(2) momentum}} (a 2$\times$2 matrix in spin space), in which the factor $\frac{g \mu_B}{2} {\bm \sigma} \times {\bf E}$ is proportional to an {\it{SU(2) vector potential}}.  In the ring geometry of Fig.~\ref{Fig4}(b), the magnitude of the electric field is constant but its direction is angle dependent -- it points radially outward, 
${\bf E} = (\lambda/L){\hat {\bf r}}$.  The last term on the RHS of (\ref{ACRing}) is a constant,
$\bar {\veps} = \frac{(g \mu_B {\bf E})^2}{8 m_e c^2}$.  The Schr\"odinger equation for an electron on the ring 
involves only the angular direction $\hat {\theta}$ along the ring, ${\bm \sigma} \times {\bf E}$
is proportional to to $\sigma_z$, so
\begin{equation} \label{Pauli}
    \frac{1}{2m_e}(p_\theta+\frac{e}{ c} \frac{\Phi_{AC}}{L} \sigma_z)^2 \psi =
    \veps(\Phi_{\text AC}) \psi, ~ \ \  \Phi_{AC} \equiv \frac{g \mu_B \lambda}{2 e}.
\end{equation}
where $p_\theta$ is the tangential momentum and the Aharonov--Casher phase $\Phi_{AC} \equiv \frac{g \mu_B \lambda}{2 e}$ has
units of magnetic flux. Thus, in this simple example, the operator
\begin{equation} \label{ACVP}
A_\theta \equiv \frac{g \mu_B \lambda}{2e L} \sigma_z  = \frac{\Phi_{AC}}{L} \sigma_z ,
\end{equation} 
can be viewed as a tangential component of a $2$$\times$$2$ matrix SU(2) vector potential. 
The similarity of (\ref{Hamiltonian}) and (\ref{Pauli}) is evident.  In Eq.~(\ref{Hamiltonian}),
$\Phi_{\text AB}/L$ is the vector potential on the ring responsible for the Aharonov--Bohm effect, whereas 
in Eq.~(\ref{Pauli}), $\Phi_{AC}\sigma_z/L$ can be interpreted as a $2$$\times$$2$ matrix
vector potential on the ring responsible for the Aharonov--Casher
effect.  This analogy is not perfect because in the latter case, the
energy eigenvalue contains a constant $\bar {\veps}=\left [\frac{e}{ c} \frac{\Phi_{AC}}{L} \right ]^2$ hence strictly
speaking, $\veps(\Phi_{AC})$ is not periodic in $\Phi_{AC}$.
However, the ratio between $\bar {\veps}$ 
and the kinetic energy at the Fermi momentum $p_F$ can be shown to be extremely small, it is legal to assume that $\bar {\veps}$ can safely be neglected.  
Then, $\Phi_{AB}/L$ and $\Phi_{AC}\sigma_z/L$ appear on the same footing as $U(1)$ and $SU(2)$ vector potentials respectively.
 
The energy spectrum can be understood in analogy with
the discussion of the spectrum of the Aharonov--Bohm system of electrons on a ring. The wave functions
for spin up electrons ($|+ \rangle \equiv |\!  \!  \uparrow \rangle$)
and spin down electrons ($|- \rangle \equiv |\!  \!  \downarrow
\rangle$) are $\psi_n^{(\pm)}(x) = L^{-1} e^{i k_n x}|\pm \rangle$,
where $k_n=\frac{2 \pi n}{L}$ with $n=0,\pm1, \pm 2, \ldots$, and the
energy eigenvalues are,
\begin{equation} \label{Eq:MSI_EnAC}
    E_n^{(\pm)} = \frac{\hbar^2}{2 m_e R^2} (n \pm \gamma)^2-{\bar
    E}=\frac{\hbar^2}{2 m_e R^2}(n^2 \pm 2 n \gamma) ~,
\end{equation}
where $\gamma = \Phi_{AC}/\Phi_0 = \frac{g \mu_B \lambda}{2 h
c}=\frac{g e \lambda}{2 \pi m_e c^2}$, and $\bar {E}=\frac{(g \mu_B
{\bf E})^2}{8 m_e c^2}=\frac{\hbar^2 \gamma^2}{2 m_e R^2}$.

\subsection{Aharonov--Casher effect and non-Abelian gauge transformations}
The example discussed above and illustrated in Fig.~\ref{Fig4}(b) is a simple particular case wherein the vector potential $\frac{g \mu_B \lambda}{2e L} \sigma_z$ involves a {\it single Pauli matrix} and a line integral between two points, e.g., $\int_{x_q}^{x_p} e^{-i \frac{e}{\hbar c} \frac{\Phi_{AC}}{L}\sigma_z}dx$, makes sense.  Generically, (as shown below), $SU(2)$ vector potentials depend on space point ${\bf r}$. The components of the $SU(2)$ vector potential along a line involve combinations of Pauli matrices such as $a_x({\bf r})\sigma_x+a_y({\bf r})\sigma_y+a_z({\bf r})\sigma_z$, where the coefficients are real numbers.  Thus, in the language of non integrable phase factors, the phase factor $\Lambda_{pq}$, Eq.~(\ref{nipf}) is an element of the $U(1)$ gauge Abelian group represented here by complex numbers on the unit circle.  On the other hand, in the Aharonov--Casher effect, we encounter the case where non-integrable phase factors require integration along a given curve connecting points $p$ and $q$ that should be path ordered, that is,  the $SU(2)$ non-integrable phase factor is,
\begin{equation} \label{nipfAC}
    \eta_{pq}=\big [\int_{{\bf r}_q}^{{\bf r}_p} e^{-i \frac{e}{\hbar c} {\bf A} \cdot d {\bf r}}
     \big ]_{\text{path ordered}} 
\end{equation} 
Gauge theories play a central role in many fields of modern physics.  Non-Abelian gauge theories were introduced into physics in a seminal paper (Yang and Mills 1954).

\subsection{Examples in condensed matter physics}
\begin{figure}[!ht]
\centering
\includegraphics[width=0.5\textwidth]{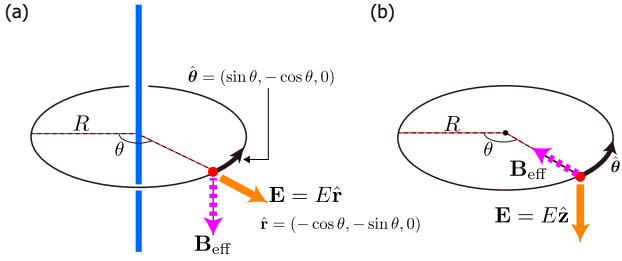}
\caption{
Two cases where an electron on a ring is subject to an electric field that is responsible for spin-orbit coupling.  (a) As in Fig.~\ref{Fig4}(b), the electron is subject to a {\it{radial electric field}} $\mathbf{E} = E\hat{\mathbf{r}}$ generated by an infinite uniformly charged wire stretched along the ring axis. In its rest frame the electron feels a constant effective magnetic field along $\hat {\bf z}$.  In (b), the electric field (generated, e.g, by a gate voltage), is perpendicular to the ring $\mathbf{E} = E\hat{\mathbf{z}}$. Note that in (a) the effective magnetic field is constant and the phase factor is calculated by an ordinary integral. However, in (b) the electric field generates a {\it{position-dependent effective magnetic field}}, and the phase factor is calculated by a path-ordered integral as in Eq.~(\ref{nipfAC}). The pertinent coupling is referred to as {\it{Rashba spin-orbit coupling}}.  Taken from Avishai {\it{et al}}. 2019. 
}
\label{Fig5}
\end{figure}

Let us consider an electron moving on a metallic ring 
of radius $R$ lying on the plane $z=0$
centered at the origin of the $x$-$y$ plane (Meijer 2002). 
Its position is specified by the polar vector 
 ${\bf r}=(R,\theta)$ [see Fig.~\ref{Fig5}(b)]. 
 \ \\

The electron is subject to an electric field ${\bf E}=E\hat{\bf z} \perp \widehat{\bm \theta}$. Introducing $R$ as the length unit enables us to rewrite the corresponding Schr\"odinger equation for the electron on the ring  in terms of dimensionless quantities. The mechanical momentum is $p_\theta=-i \frac{d}{d \theta}$.  Within the SU(2) formulation of the Pauli equation (Pauli 1927, Anandan 1989, Frohlich 1993), the SU(2) vector potential in the ring geometry  takes the form, 
\vspace{-0.in}
\begin{equation} \label{2a}
{\bm A}(\theta)=\beta(\theta) \widehat{\bf E}(\theta)\times {\bm \sigma},
 \vspace{-0.03in}
\end{equation}
where the real parameter $\beta(\theta)$ specifies the strength of the local spin-orbit coupling (Meir {\it{et al}}. 1989, Meir {\it{et al}}. 1990, Entin Wohlman {\it{et al}} 1992) [compare with Eq.(\ref{ACVP}) applicable for case Fig.~\ref{Fig5}(b)].  For simplicity it is assumed that $\beta$ is independent of $\theta$. For example, if the Rashba Hamiltonian [relevant for the present discussion based on figure \ref{Fig5}(b)], is written as 
$H_{\text{R}} =-\alpha_{\text{R}}   ({\bm \sigma} \times {\bf p}) \cdot \hat {\bf E}$   
(where $\alpha_{\text{R}}=-\tfrac{1}{2 m c} g \mu_{\text B}\vert {\bf E} \vert$ is the Rashba SOC strength parameter), hence $\beta=2mR \alpha_{\text{R}}/\hbar^2$. 

In the present ring geometry, ${\bf k}$ is parallel to the tangential unit vector $\widehat{\bm \theta}$ 
and we are concerned only with the tangential component of 
$ {\bm A}(\theta)$, that is an element of the su(2)-algebra, 
\begin{equation} \label{calA} 
 {\mathcal A}(\theta)={\bm A}(\theta)\cdot \widehat{\bm \theta} 
 =\beta [ \widehat{\bm \theta} \times \widehat{\mathbf{E}} ]   \cdot {\bm \sigma} 
 \equiv \beta \, \widehat{\bf n}\cdot {\bm \sigma} 
 \;  \left[ \in \mbox{su(2)}  \right].
 \end{equation} 
 Classically, $ \widehat{\bm \theta} \parallel {\bf v}$ (where ${\bf v}$ is the electron velocity)
so that  $\widehat{\bf n}$ points along ${\bf E} \times {\bf v}$ that is the direction of the
effective magnetic field ${\bf B}_{\text{eff}}$
felt by the electron in its rest-frame. Quantum mechanically, $\widehat{\bf n}$ encodes the nature of the spin-orbit coupling. In our example for electrons on a ring subject to a perpendicular  electric field, 
$\widehat{\bf n} = \widehat{\bm \theta} \times \widehat{\mathbf{E}}$. In  two-dimensional semiconductors it is possible to have also an intrinsic spin-orbit coupling (Engel {\it{et al}}. 2006). In one simplified form (Dresselhaus 1955), derived from the Kane model of ${\bf k} \cdot {\bf p}$ perturbation theory (Kane 1957), $\widehat{\bf n}$ is defined through $\widehat {\bf n} \cdot {\bm \sigma} = \cos \theta \, \sigma_x - \sin \theta \, \sigma_y$ (Engel {\it{et al}}. 2006, Eq. (8)).
With the unit vector $\widehat {\bf n}$, the SU(2)-invariant Schr\"odinger equation (in dimensionless units) for  the spinor $\psi(\theta)=\binom{\psi_\ua(\theta)}{\psi_\da(\theta)}$ is written as:
\begin{equation} \label{dlse}
\left[-i \frac {d}{d\theta}+ \beta \, \widehat{\bf n}(\theta) \cdot {\bm \sigma} \right]^2 \psi(\theta)=\veps \psi(\theta) \, ,
\end{equation} 
where $\veps = \frac{2 m R^2}{\hbar^2} {\cal E} \equiv k^2$, with ${\cal E}$ being the electron energy in physical units. 

Now we are in a position to express the non-integrable phase factor. 
If the SU(2)  vector potential  is changed
from $\theta$ to $\theta+ d\theta$  
along an infinitesimal directed arc $d{\bm \ell} = R \hat{\bm \theta} d \theta$, it gains an  SU(2) {\it{matrix-valued}} phase factor according to
\begin{equation} \label{2}
{\bf A}(\theta+d \theta)=e^{i \beta \, \widehat{\bf n}(\theta) \cdot {\bm \sigma} d \theta} {\bf A}(\theta).
\end{equation}    
If the spin-orbit coupling strength $\beta$ and/or the unit vector $\hat {\bf n}$ depend on the position $\theta$, two phase factors do not commute.  Thus, the SU(2) non integrable phase factor accumulated from $\theta=0$ up to a finite angle $\theta$ is given by the following path-ordered integral, or equivalently, an ordered product of infinitesimal phase factors, (Wu and Yang 1985, Anandan 89). The {\it{non-abelian}} gauge transformation $\psi(\theta) \to \xi(\theta)$ that eliminates the vector potential is, 
\begin{equation} 
\begin{split}
\psi(\theta) &= \mathcal{P}  \exp \left\{ \int_{0}^{\theta} i \beta \, 
\widehat{\bf n} (\theta^{\prime}) {\cdot} {\bm \sigma} d \theta^{\prime}  \right\} \xi(\theta)  \\
&= \lim_{N \to \infty} \left\{ \prod_{\substack{j=1\\ \longleftarrow }}^{N}
\be^{ i \beta \, \widehat{\bf n}(j \Delta \theta) {\cdot} {\bm \sigma} 
\Delta \theta  } \right \} \xi(\theta) \\
&\equiv \Lambda_{\text{AC}}[\theta ;\beta] \xi(\theta),
\end{split}
\label{eqn:Wilson-line-discretized}
\end{equation}
where $ \mathcal{P}$ is path ordering, $\Delta \theta =\theta/N$ and the arrow $\longleftarrow$ means that matrices are multiplied from the right to the left.  The $2$$\times$$2$ matrix function $\Lambda_{\text{AC}}[\theta ;\beta]$ satisfies a first order differential equation that can be solved analytically (Avishai {\it{et al}}. 2019).  The function $\xi(\theta)$ then satisfies Eq.~(\ref{dlse}) in the absence of the vector potential. The procedure of multiplying phase factors as in Eq.~(\ref{eqn:Wilson-line-discretized}) and defining $\lambda_{\text{AC}}$ by Eq.~(\ref{ACPF}) can be extended straightforwardly to any continuous closed curve. 

Since the gauge transformation $\Lambda_{\text{AC}}[\theta ; \beta]$ is an SU(2) matrix, it can be written as $\Lambda_{\text{AC}}[\theta ; \beta] =\be^{i \lambda(\theta, \beta) \widehat{{\bf b}}[\theta, \beta] \cdot {\bm \sigma}}$, where $\lambda(\theta, \beta)$ and the two dimensional unit vector $\widehat{{\bf b}}[\theta,\beta]$ should be calculated within a given spin-orbit coupling scheme (see example the analysis of the scattering in the square geometry below).   The Aharonov--Casher phase $\lambda_{\text AC}$ is here defined as a single number in terms the SU(2) phase acquired along the entire circle:
\begin{eqnarray}  
&&F_{\text{AC}}[2 \pi; \beta] = \mathcal{P}  \exp \left\{ \oint
 i \beta \, \widehat{\bf n} (\theta^{\prime}) {\cdot} {\bm \sigma} d \theta^{\prime}  \right\} \\ \nonumber
&& \, \, \equiv  e^{i \lambda_{\text{AC}} \widehat{\bf b} \cdot {\bm \sigma}}
 = \underbrace{\cos \lambda_{\text{AC}} \, \mathbf{1}_{2 \times 2}}_{\text{non-vanishing trace}}
 +\underbrace{i \sin \lambda_{\text{AC}} \,
 \widehat{\bf b} \cdot{\bm \sigma}}_{\text{traceless}} \; .
\label{ACPF}
 \end{eqnarray}
Consequently, three real parameters, i.e., the angle $\lambda_{\text{AC}}$ and the two-component unit vector $\widehat{\bf b}$  determine the Aharonov--Casher phase factor unambiguously. The wave function $\psi(\theta) = \Lambda_{\text{AC}}[\theta; \beta] \xi(\theta)$ should be periodic, that is, $\psi(2 \pi)=\psi(0)$, and this equality determines the energy eigenvalues.

In analogy with persistent charge current discussed in relation to the Aharonhov-Bohm effect we encounter here the possible occurrence of {\it spin current} (Sun and Xie, Niu 2006). It is formally  defined as $J_S(\theta)= \tfrac{1}{2}\psi^\dagger (\theta) \{{\bf v}, {\bm \sigma}\}\psi(\theta)$, where ${\bf v}=\left[-i \frac {d}{d\theta}+ \beta \, \widehat{\bf n}(\theta) \cdot {\bm \sigma} \right]$ is the velocity operator. Unlike charge current, spin current is time reversal invariant and, (generically) space dependent. 

\subsubsection{Aharonov--Casher interferometer in a perpendicular electric field}

\begin{figure}[!ht]
\centering
\includegraphics[width=0.45\textwidth]{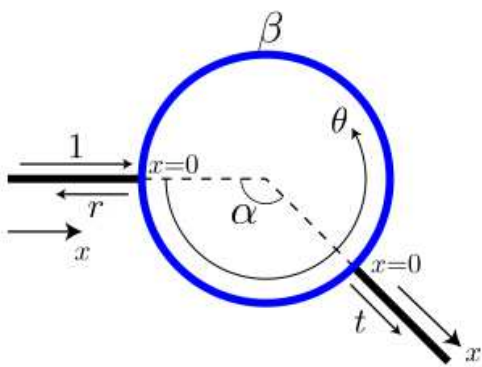}
\caption{ Aharonov--Casher  Interferometer (taken from Avishai {\it{et al}}. 2019): Similar to Fig.~\ref{Fig3},
two wires at different potentials $V_L  \gtrapprox  V_R$ are connected to a metallic ring. 
Electrons approaching  the ring from the left at polar angle $\theta=0$ are partially reflected back to the left wire and partially transmitted to the right wire  at polar angle $\theta=\alpha$ (reflection matrix $r$ and transmission matrix $t$).  The ring is subject to a perpendicular electric field, so that the Rashba spin-orbit mechanism generates an effective magnetic field along the radial direction $\hat {\bf n}(\theta)=(\cos \theta, \sin \theta,0)$ [Fig.~\ref{Fig5}(b)].  The electric field is {\it{homogeneous and acts only on the circular area confined by the ring}}. The local phase factor for {\bf any} polar angle $\theta$ is $e^{i \beta \, \hat {\bf n}(\theta) \cdot {\bm \sigma}}$.  Taken from (Avishai {\it et al}. 2019), where an analysis of Aharonov--Casher interferometer subject to some version of Dresselhaus spin-orbit coupling is also analyzed. }
\label{Fig6}
\end{figure}

Having studied the physics of an Aharonov--Bohm interferometer, it is expected, following an experiment  by (Borunda {\it{et al}}. 2008), that there is an analogous system for the Aharonov--Casher effect. The setup of a simple example of Aharonov--Casher interferometer is illustrated in Fig.~\ref{Fig6}.  To write down the Schr\"{o}dinger equation, we use, as before, dimensionless quantities.  The electron wave function at point $\bfr$ (any point on the ring or on either wire) for a boundary condition that an initial incoming electron has spin projection $\sigma$ is expressed as a spinor: 
$\Psi_\sigma(\bfr) = \binom {\psi_{\ua \sigma} (\bfr)}{ \psi_{\da \sigma} (\bfr) }$.  The two spinors for the spin projections $\sigma = \uparrow, \downarrow$ of the incoming electron can be arranged into a $2$$\times$$2$ wave function matrix,
$$ {\bm \Psi}(\bfr) = \begin{pmatrix} \psi_{\ua \ua} (\bfr) &  \psi_{\ua \da} (\bfr) \\  \psi_{\da \ua} (\bfr) &  \psi_{\da \da} (\bfr) \end{pmatrix},$$
so we have,
\begin{widetext}
\begin{equation} \label{SE}
\begin{cases} -\frac{d^2}{d x^2} {\bm \Psi}(x) = k^2 {\bm \Psi}(x), \quad  (\text{when the electron is on left or right wire}) \\
\left[  -i \frac{d}{d \theta}-\beta \hat{\bf n}(\theta) \cdot {\bm \sigma} \right]^2 {\bm \Psi}(\theta)=k^2{\bm \Psi}(\theta) 
\quad  (\text{when the electron is on the ring}) \; .
\end{cases}
\end{equation}
\end{widetext}
We consider an electron with spin projection $\sigma = \ua, \da$ approaching the ring from the left wire 
at energy $\veps=k^2$. It is partly transmitted into the right wire with spin projection $\mu = \ua, \da$ 
and partly reflected back into the left wire with $\mu= \ua, \da$. The corresponding matrix elements of the $2 \times 2$ transmission and reflection matrices are $t_{\mu \sigma}, r_{\mu \sigma}$, explicitly, 
$$ t=\begin{pmatrix} t_{\ua\ua} & t_{\ua\da}\\ t_{\da\ua} & t_{\da\da} \end{pmatrix} \ \  
 r=\begin{pmatrix} r_{\ua\ua} & r_{\ua\da}\\ r_{\da\ua} & r_{\da\da} \end{pmatrix}.$$  
The technique for calculating $t$ and $r$ is similar to that used in the Ahronov--Bohm interferometer, but is slightly more complicated. The conductance through the interferometer is given by the Landauer formula (Landauer 1957), $g=\frac{e^2}{h} \mbox{Tr}[t^\dagger t]$.  

\begin{figure}[htb]
\centering
\subfigure[]
{\includegraphics[width=0.45\textwidth]{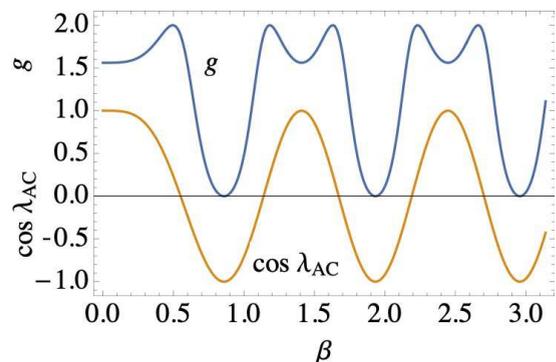}}
 \subfigure[]
 {\includegraphics[width=0.45\textwidth] {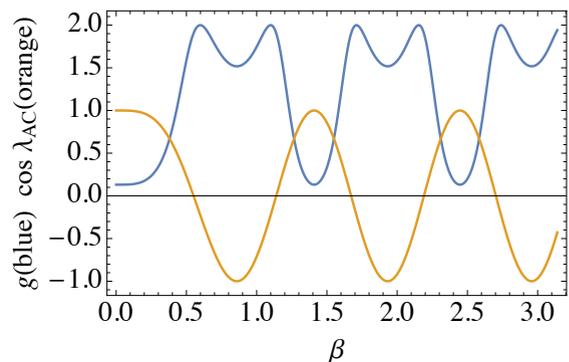}} 
\caption{Plots of dimensionless conductance $g$ (blue line) and cosine of the Aharonov--Casher phase $\cos \lambda_{\text{AC}}$ (orange line) 
as function of spin-orbit strength $\beta$
for the ring interferometer schematically depicted in Fig.\ref{Fig6}. The wave number is set to $kR=1.25$, and 
(a)  $\alpha=\pi$, and (b) $\alpha=1.7$.  
As a property of the closed loop,  $\lambda_{\text{AC}}$ is independent of $\alpha$. The conductance,
that also encodes the geometric properties of the interferometer naturally depends on $\alpha$, but in both cases it is periodic in the Aharonov--Casher phase $\lambda_{\text AC}$. Taken from Avishai {\it{et al}}. 2019.}
\label{Fig7}
\end{figure}

Figure~\ref{Fig7} plots the cosine of the Aharonov--Casher phase $\lambda_{\text AC}$ and the conductance $g$ (in units of $\frac{e^2}{h}$) as a function of the spin-orbit strength $\beta$ for two values of the right splitter angle $\alpha$. 

\subsubsection{Conductance and Polarization in transmission through a square interferometer}

It was shown in the discussion of Eq.(\ref{ACPF}) that the SU(2) phase factor $e^{i \lambda_{\text AC}\hat{\bf b}\cdot {\bm \sigma}}$ can be written as a sum of two $2 \times 2$ matrices, one has non-vanishing trace and one is traceless. Both depend on the Aharonov--Casher phase $\lambda_{\text AC}$, but the role of the trace-less term (in particular the unit vector $\hat {\bf b}$) was not explained. Using a relatively simple tight-binding model that can be solved analytically, it will  be shown that this term is responsible to electron polarization. Note that spin-orbit coupling obeys time reversal invariance but the specification of scattering boundary condition destroys it, so that polarization is feasible (Entin-Wohlman et al 2010).

\begin{figure}[htb]
\begin{center}
{\includegraphics[width=0.5\textwidth]{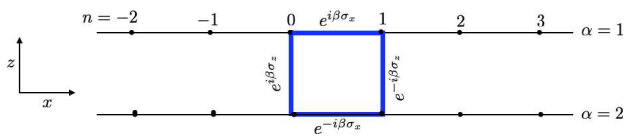}}
\caption{Two-channel tight-binding model for electron scattering from a square region in the $x$-$z$ plane  (shown by bold blue lines) subject to a uniform electric field $E\hat {\bf y}$ acting only on the square. The corresponding SU(2) hopping matrix elements are given by $e^{\pm  i \beta \sigma_x}$ along the horizontal links and $e^{\mp  i \beta \sigma_z}$ along the vertical links. (Taken from Avishai {\it et al}. 2019).
 }
\label{Fig8}
\end{center}
\end{figure}

It can be shown (Avishai {\it{et al}}. 2019) that in two-port systems like the Aharonov--Casher interferometer discussed above, there is no polarization in the outgoing lead (assuming that the incoming beam is not polarized). Here we show that in a multiport system, it is possible to get a polarized  outgoing beam due Rashba spin-orbit coupling.  Consider, as in Fig. \ref{Fig8}, a tight-binding model of non-interacting electrons moving along two wires and scattered from a region that carries an SU(2) flux where the phase factors in the product do not commute and depend on the spin-orbit strength. The system is laid on the $x-z$ plane and consists of two chains (metallic wires) numbered $\alpha=1,2$ with each chain consisting of sites $-\infty < n < \infty$, which are then connected to each other by the two rungs of the square at $n=0$ and $n=1$ (see Fig.~\ref{Fig8}).   A uniform electric field acts along $\hat {\bf y}$ upon the square thereby leads to a Rashba spin-orbit coupling along the links of the square.  This tight-binding model is treated within the formalism of second quantization. The creation and annihilation operators for the spin-projection $\sigma=\uparrow,\downarrow$ are indexed as $c^\dagger_{\alpha, n, \sigma}$ and $c_{\alpha, n, \sigma}$ respectively, and the spin-orbit interaction is active only on the four links forming the square (shown by the bold lines in Fig.~\ref{Fig8}). The Hamiltonian is written as $H$=$H_0+H_1$ with 
\begin{equation}
\begin{split}
& H_0=-t \sum_{\alpha=1,2} \sum_{n \le 0}{\bf c}^\dagger_{\alpha, -(n+1)} {\bf c}_{\alpha, -n} \\
& \qquad - t  \sum_{\alpha=1,2}\sum_{n \ge 1}{\bf c}^\dagger_{\alpha, (n+1)} {\bf c}_{\alpha, n}+ \text{h.c.} ,  \\
& H_1=\sum_{n=0,1}{\bf c}^\dagger_{1,n} e^{- i \beta \sigma_x}{\bf c}_{2,n} 
+\sum_{\alpha=1,2}{\bf c}^\dagger_{\alpha,0} e^{i \beta \sigma_z}{\bf c}_{\alpha,1}+  \text{h.c.},
\end{split}
\label{H}
\end{equation} 
where ${\bf c}^\dagger_{\alpha,n} \equiv (c^\dagger_{\alpha,n, \ua},c^\dagger_{\alpha,n, \da})$.

The solution of the scattering problem yields the $4$$\times$$4$ (2 for spin and 2 for channel) transmission and reflection  matrices $t$ and $r$ whose matrix elements $t_{\alpha' \sigma';\alpha \sigma}$ and $r_{\alpha^{\prime} \sigma^{\prime}; ;\alpha \sigma}$ give the transmission and reflection amplitudes for scattering of a spin-$\sigma$ electron impinging from channel $\alpha$ to one in the channel $\alpha^{\prime}$ with spin projection $\sigma^{\prime}$. They depend on the SOC strength, and the wave number $k$ that determines the energy $\veps= - 2 t \cos k$.  On the other hand, the Aharonov--Casher phase $\lambda_{\mathrm{AC}}$, is a property of the closed loop irrespective of the scattering energy.  In the present case, it is given by calculating the product of the four matrices shown in Fig.~\ref{Fig8}, and using Eq.~(\ref{ACPF}), giving
\begin{equation} 
\cos \lambda_{\mathrm{AC}}(\beta) = 1-2 \sin^4\beta. 
\label{PhaseSquare1}
\end{equation} 
Before presenting results related to the scattering problem, it is worth while 
to check the eigenvalue problem of the a system of an electron hopping 
on an isolated square (without wires; the square shown by bold lines in Fig.~\ref{Fig8}) as a closed system, 
especially whether they depend only on the Aharonov--Casher phase. 
The tight-binding Hamiltonian assumes the following form
\begin{equation} \label{Hsquare}
H_{\square}=\begin{pmatrix} 
0&e^{i \beta \sigma_z}&e^{-i \beta \sigma_x}&0\\
e^{-i \beta \sigma_z}&0&0&e^{-i \beta \sigma_x}\\
e^{i \beta \sigma_x}&0&0&e^{i \beta \sigma_z}\\
0&e^{i \beta \sigma_x}&e^{-i \beta \sigma_z}&0 \end{pmatrix}  \;, 
\end{equation} 
where each entry is a $2 \times 2$ matrix acting on the spin space at each site of the square. Simple calculations 
shows the there are four different eigenvalues each of which is two-fold (Kramers) degenerate: 
\begin{equation}
E_\square=\pm 2 \cos \tfrac{ \lambda_{\mathrm {AC}}(\beta)}{4} , \ \ 
\pm 2 \sin \tfrac{ \lambda_{\mathrm {AC}}(\beta)}{4}. 
\end{equation}
In any case, the eigenvalues depends on $\beta$ only through $\lambda_{\text{AC}}(\beta)$ as defined in Eq.~(\ref{PhaseSquare1}). 

The solution of the scattering problem yields the $4$$\times$$4$ (2 for spin and 2 for channel) transmission and reflection  matrices $t$ and $r$ whose matrix elements $t_{\alpha' \sigma';\alpha \sigma}$ and $r_{\alpha^{\prime} \sigma^{\prime};\alpha \sigma}$ give the transmission and reflection amplitudes for scattering of a spin-$\sigma$ electron impinging from channel $\alpha$ to one in the channel $\alpha^{\prime}$ with spin projection $\sigma^{\prime}$. They depend on the spin-orbit coupling strength $\beta$, and the wave number $k$ that determines the energy $\veps= - 2 t \cos k$.  
Going back to the scattering problem, we will now inspect the relation between the ACP and a couple of experimental observables.   Specifically, we  are interested in the conductance $g$ and the transmitted polarization vector ${\bf P}$, defined as, 
\begin{equation} \label{gP}
g(k;\beta) 
=\mbox{Tr} [t^\dagger t ], \ \ {\bf P}=\frac{\mbox{Tr} [t^\dagger {\bm \Sigma} t ]}{g} , \ \ {\bm \Sigma}={\bf 1}_{2 \times 2} \otimes {\bm \sigma}.
\end{equation} 
The following expressions for the conductance 
$g$ and the transmitted polarization $P_y$ are as follows:
\begin{widetext}
\begin{equation}
 \label{gPy}
 \begin{split}
& g=\frac{16 (-5+4 \cos 2k(1-\cos \lambda_{\text{AC}})\sin^2 k}
{-17+8 \cos 2k +8(\cos 4k-4 \cos 2k)(1-\cos \lambda_{\text{AC}})-16(1-\cos \lambda_{\text{AC}})^2}, \\
 & P_y=\frac{2 \sin 2k(1-\cos \lambda_{\text{AC}})}{5-2 \cos 2k(1-\cos \lambda_{\text{AC}})}, \ \ P_x=P_z=0.  
\end{split}
\end{equation}
\end{widetext}
When the spin-orbit coupling strength $\beta \to 0$ then $\lambda_{\text AC} \to 0$. The conductance remains finite but the polarization vanishes (as it must). But when $\lambda_{\text AC} \to 0$, the traceless part of the Aharonov--Casher phase factor in Eq.~(\ref{ACPF}) tends to zero as well. Hence, this part of the phase factor is related (albeit in a complex way) to an experimental quantity.  

\section{Summary}
Aharonov--Bohm and Aharonov--Casher effects substantiate the fundamental role of quantum mechanics as a solid theory of Nature (Richter 2012). In addition, they expose its beauty and peculiarities, and, at the same time, point out the possible occurrence of its puzzles. The Aharonov--Bohm effect constitutes a simple manifestation of Abelian gauge theory, while  the Aharonov--Casher effect constitutes a simple manifestation of non-Abelian gauge theory. These two effects are published several decades ago, and  remain among the building blocks of quantum mechanics ever since. The question of duality between the Aharonov--Casher and Aharonov--Bohm effects is raised by (Hagen 1990b, Bogachek and Landman 1994, Borunda 2008, Rohrlich 2010). A gedanken experiment relating the two effects is analyzed in Ref. (Avishai and Luck 2009). 


Both effects stress the essential role of phase of the wave functions (Berry 1984).  The Aharonov and Bohm effect involves the motions of charged particles (e.g, electrons) around but away from a region in which the magnetic field is finite. In the Aharonov and Casher effect (Aharonov and Casher 1984), an electrical field is considered and its effect on particles with non-zero magnetic moment such as electrons (but also neutrons as they need not be charged), is analyzed.  Both of these effects have direct relevance to condensed matter systems.  Here we described these two effects and review their relevance in condensed matter (mainly in mesoscopic systems).

In this short review, some aspects of the relevance of these effects to condensed matter physics are illuminated.  In particular,  we focus on the physics of mesoscopic systems, for which numerous experiments confirm the two effects.  For the Aharonov--Bohm effect, examples of interference around a tube carrying magnetic flux,  electrons on the ring (spectrum and persistent current), and an interferometer are analyzed.  For the Aharonov--Casher effect, examples of an interferometer, and an analysis of transmission and polarization in a two channel system are presented. \\
 It is the author's hope that the broad list of references should enable the readers to become acquainted with other aspects of these effects in condensed matter physics. \\
 \ \\
 \underline{Acknowledgement}: The authors thank Tapash Chakraborty for suggesting this project, and Lev Vaidman for valuable comments. 

\newpage
\begin{widetext}
\begin{center} 
REFERENCES
\end{center}

\reversemarginpar\marginnote{\bf A \ \   }
Aharonov and Bohm 1959.\\
\noindent
Aharonov, Y. Bohm, D. ``Significance of electromagnetic potentials in quantum theory". Phys. Rev. 115 (3) 485-491.  \\
\noindent
Aharonov and Bohm 1961.\\
Aharonov, Y; Bohm, D. ``Further Considerations on Electromagnetic Potentials in the Quantum Theory". Physical Review. 123 (4): 1511-1524. \\
\noindent
Avishai {\it{et al}}. 1972\\
Avishai Y, Ekstein, H and   Moyal, J. E. ``Is the Maxwell Field local?" J. Math. Phys.  13, 1139.\\
\noindent
Altshuler 1981\\
Altshuler, B. L. Aronov, A.G. Spivak,   B. Z. ``The Aharonov--Bohm effect in disordered conductors'', 
JETP Lett. (Engl. Transl.) 33, 94.\\
\noindent 
 Aharonov and Casher 1984.\\
Aharonov, Y. Casher, A. ``Topological quantum effects for neutral particles",
 Phys. Rev. Lett. {\bf 53} (4): 319-321.\\ 
 \noindent
 Aharonov and Anandan 1987\\
 Aharonov, Y.  Anandan, J. ``Phase change during a cyclic quantum evolution", 
 Phys. Rev. Lett. {\bf 58}, 1593.\\
 \noindent Avishai and Band 1987\\
 Avishai, Y. and Band, Y. ``Quantum scattering determination of magneto-conductance for 2D systems'', 
 Phys. Rev. Lett. {\bf 58}, 2251.\\
 \noindent
 Aharonov {\it{et al}}. 1988\\
 Aharonov, Y. Pearle, P. Vaidman, L. ``Comment on Proposed Aharonov--Casher effect: Another example of an Aharonov--Bohm effect arising from a classical lag'', 
 Phys. Rev. A{\bf 37}, 4052-4055.\\
  \noindent
 Anandan 1989\\ 
 Anandan J. ``Electromagnetic effects in the quantum interference of dipoles'', 
 Phys. Lett. A {\bf 138}, 347.\\
 \noindent
 Aronov 1993\\
 Aronov, A.G.  and  Lyanda-Geller, Y. ``Spin-orbit Berry phase in conducting rings'', 
 Phys. Rev. Lett. {\bf 70} 343.\\
 \noindent
   Avishai et.  al. 1993\\
 Avishai, Y. Hatsugai, Y. and Kohmoto M.  ``Persistent Currents and edge states in a magnetic field"
Phys. Rev. B47,  9501.\\
\noindent
Avishai, Y and Kohmoto, M. 1993\\
``Quantized Persistent Currents in Quantum Dot at Strong Magnetic Fields" 
Phys. Rev. Lett. 71, 279.\\
 \noindent
Altland et al 1996\\
Altland A, Gefen Y, and Montambaux G. ``What is the Thouless Energy for Ballistic Systems?", 
Phys. Rev. Lett. {\bf 76}, 1130. \\
 \noindent
Aharony {\it{et al}}. 2002\\
Aharony A, Entin-Wohlman O, Halperin B. I. and Imry Y.
``Phase measurement in the mesoscopic Aharonov--Bohm interferometer"
Phys. Rev. B66, 115311. \\ 
\noindent 
Aharony {\it{et al}}. 2003\\
Aharony A, Entin-Wohlman O, and Imry Y.
``Measuring the transmission of a quantum dot using Aharonov--Bohm interferometers'',
J. of the Physical Society of Japan 72, Supp. A, 112 (2003) (cond-mat/0208068).\\
\noindent
Aharony {\it{et al}}. 2005\\
Aharony A, Entin-Wohlman O, and Imry Y.
``Phase measurements in open and closed Aharonov--Bohm interferometers'',
Physica E 28, 283 (2005) (cond-mat/0412027)\\
\noindent
Aharony {\it{et al}}. 2006\\
A. Aharony, O. Entin-Wohlman, T. Otsuka, S. Katsumoto, H. Aikawa, and K. Kobayashi
``Breakdown of phase rigidity and variations of the Fano effect in closed Aharonov--Bohm interferometers'', Phys. Rev. B 73.\\
\noindent
Akkermans and Montambaux 2007\\
Akkermans E. and Montambaux G.
``Mesoscopic Physics of electrons and photons", Cambridge University Press.\\
\noindent
 Avishai, and Luck 2009\\
Avishai, Y. and Luck.  J. M. ``Magnetization of two coupled rings",
Journal of Physics A  (Math. Gen.), 42, 175301.\\
\noindent
Aharony {\it{et al}}. 2014\\
Aharony, A. Takada, S. Entin-Wohlman, O. Yamamoto, M. and Tarucha, S. 
``Aharonov--Bohm interferometry with a tunnel-coupled wire'',
New J. Phys. 16, 083015. \\
\noindent
Atsushi et. al. 2014\\
Noguchi, A.  Shikano, Y. Toyoda, K. and Urabe, S. 
``Aharonov-Bohm effect in the tunneling of a quantum rotor in a linear Paul trap", Nat. Commun. 5, 3868.
\noindent
Avishai and Band 2017\\
Avishai Y and Band Y. 
``Simple spin-orbit based devices for electron spin polarization'', Phys. Rev. B 95, 104429.\\
\noindent
Aharony {\it{et al}}. 2019\\
Aharony, A. Entin-Wohlman, O.  Tzarfati, L. H.  Hevroni, R. Karpovski, M. Shelukhina, V. Umansky, V. and  Palevski A.
``Possible observation of Berry phase in Aharonov--Bohm rings of InGaAs."
Solid State Electronics 155, 117.\\
\noindent
Avishai {\it{et al}}. 2019\\
Avishai Y, Totsuka K, and Nagaosa N. 
``Aharonov--Casher Phase Factor in Mesoscopic Systems'' ,
J. Phys. Soc. Jpn., 88, No.8, Article ID: 084705."\\
\noindent
\reversemarginpar\marginnote{\bf B \ \ \ \ }Byers and Yang 1961\\
Byers, N. and Yang, C. N, ``Theoretical Considerations Concerning Quantized Magnetic Flux in Superconducting Cylinders", Phys. Rev. Lett. 7, 46.\\
\noindent
Bloch 1970\\
 Bloch F, ``Josephson Effect in a 
Superconducting Ring", Phys. Rev. B2, 109. \\
\noindent
B\"uttiker {\it{et al}}. 1983\\
B\"uttiker, M., Imry, Y. and Landauer, R. ``Josephson behavior in small normal one-dimensional rings", Phys. Lett. 96, 365.\\
\noindent
(Berry 1984) \\
Berry, M. V. ``Quantal phase factors accompanying adiabatic changes, Proc. R. Soc. London, Ser. A 392, 45.
\noindent
B\"uttiker, {\it{et al}}., 1985, \\
B\"uttiker, M. Imry, Y., Landauer, R. and Pinhas, S. 
``Generalized many-channel conductance formula with application to small rings", Phys. Rev. B 31, 6207.\\
\noindent
B\"uttiker, 1985\\
B\"uttiker M. ``Small normal-metal loop coupled to an electron reservoir", Phys. Rev. B 32, 1846.\\
 \noindent
B\"uttiker 1986\\ 
B\"uttiker, M. ``Four-Terminal Phase-Coherent Conductance", Phys. Rev. Lett. 57, 1761.\\
\noindent
B\"uttiker 1988\\
B\"uttiker, M. ``Symmetry of electrical conduction", IBM J. Res. Dev. 32, 317.\\
\noindent
B\"uttiker 1990\\
B\"uttiker, M. ``Scattering theory of thermal and excess noise in open conductors", Phys. Rev. Lett. 65, 2901.\\
\noindent
B\"uttiket 1992\\
B\"uttiker, M. ``Flux-sensitive correlations of mutually incoherent quantum channels", Phys. Rev. Lett. 68, 843.\\
Bogachek and Landman 1994. \\
 Bogachek E. N. and Landman U. ``Aharonov-Bohm and Aharonov--Casher tunneling effects and edge states in double-barrier structures". 
 Phys. Rev. B. 50 (4): 2678-2680.\\
 \noindent
Berkovits and Avishai 1995\\
Berkovits, R. and Avishai, Y. ``Significant Interaction-Induced Enhancement of Persistent Currents in 2D Disordered Cylinders", Europhys. Lett. 29, 475.\\
 \noindent
 Bachtold et al 1999\\
Bachtold A., Strunk C., Salvetat J-P , Bonard J-M, L. Nussbaumer F., T., Sch\''onenberger C.,
``Aharonov-Bohm oscillations in carbon nanotubes". Nature {\bf 397} (6721): 673.\\
\noindent
Berkovits and Avishai 1996\\
Berkovits, R. and Avishai, Y.  ``Interacting electrons in disordered potentials: Conductance versus persistent currents", Phys. Rev. Lett. 76, 261.\\
\noindent
Bergsten 2006\\
 Bergsten, T. Kobayashi, T. Sekine, Y. Nitta, J. ``Experimental demonstration of the time reversal Aharonov--Casher effect", Phys. Rev. Lett. 97 196803.\\
\noindent
Borunda 2008 \\
M. F. Borunda, Xin Liu, Alexey A. Kovalev, Xiong-Jun Liu, T. Jungwirth, and Jairo Sinova
``Aharonov--Casher and spin Hall effects in two-dimensional mesoscopic ring structures with strong spin-orbit interaction'', Phys. Rev. B. 78 (24): 245315.\\
\noindent
Batelaan and Tonomura 2009\\
Batelaan, H. and Tonomura, A. ``The Aharonov-Bohm effects: Variations on a Subtle Theme'', Physics Today {\bf 62} (9): 38-43. \\
Band and Avishai 2012\\
Band, Y. Avishai, Y. ``Quantum Mechanics with Application to 
Nanotechnology and Information Science'', 
(Elsevier).  \\
\noindent 
\reversemarginpar\marginnote{\bf C \ \ \ \ } Chambers 1960\\
Chambers, R.G. ``Shift of an Electron Interference Pattern by Enclosed Magnetic Flux". Phys. Rev. Lett. 5 (1): 3-5.\\
\noindent
Cimmino et al, 1989\\
A. Cimmino, G. I. Opat; A. G. Klein; H. Kaiser; S. A. Werner; M. Arif; R. Clothier (1989). ``Observation of the topological Aharonov--Casher phase shift by neutron interferometry'', Phys. Rev. Lett. 63 (4): 380-383. \\
\noindent
Cheung {\it{et al}}. 1989\\
Cheung, H. F., Riedel, E. K. and Gefen, Y.
``Persistent currents in mesoscopic rings and cylinders'',
Phys. Rev. Lett. 62 (5), 587.\\
\noindent
Cernicchiaro 1997\\
Cernicchiaro, G. Martin, T. Hasselbach, K.  Mailly, D.  Benoit, A. ``Channel
interference in a quasi-ballistic Aharonov--Bohm experiment'', Phys. Rev. Lett. 79, 273.\\
\noindent
\reversemarginpar\marginnote{\bf D \ \ \ \ } Dresselhaus 1955\\
Dresselhaus, G. ``Spin--Orbit Coupling Effects in Zinc Blende Structures". Phys. Rev. 100 (2): 580\\
\noindent
Dowker 1967\\
Dowker, J. S. ``A gravitational Aharonov--Bohm effect". Il Nuovo Cimento B. Series 10. 52 (1): 129-135. \\
\noindent
\reversemarginpar\marginnote{\bf E \ \ \ \ }Edwards 1972\\
Edwards J. T, and Thouless, D, J, ``Numerical studies of localization in disordered systems'', J. Phys. C: Solid State Phys. 5, 807.\\
\noindent
Entin-Wohlman and Gefen 1989\\
Entin-Wohlman O, Gefen Y.
``Persistent currents in two-dimensional metallic cylinders: a linear response analysis'',
Europhysics Letters 8, (5), 477. \\
\noindent
Entin-Wohlman {\it{et al}}. 1992\\
Entin-Wohlman O, Gefen, Y,  Meir, Y Oreg Y,
"Effects of spin-orbit scattering in mesoscopic rings: Canonical-versus grand-canonical-ensemble averaging'',
Phys. Rev. B 45 (20), 11890\\
\noindent
Elion {\it{et al}}. 1993\\
 W. J. Elion; J. J. Wachters; L. L. Sohn; J. D. Mooij (1993). ``Observation of the Aharonov--Casher effect for vortices in Josephson-junction arrays'', Phys. Rev. Lett. 71 (14): 2311-2314.\\
\noindent
Entin-Wohlman {\it{et al}}. 2003\\
Entin-Wohlman O,  Imry Y, and Aharony A,
``Persistent currents in interacting Aharonov--Bohm interferometers'',
and their enhancement by acoustic radiation"
Phys. Rev. Lett. 91, 046802 (cond-mat/0302146)\\
\noindent 
Entin-Wohlman {\it{et al}}. (2004)\\
Entin-Wohlman O,  Imry Y, and Aharony A, ``Effects of external radiation on biased Aharonov--Bohm rings'',
Phys. Rev. B 70, 075301 (2004) (cond-mat/0311609)\\
\noindent
Engel et al 2006\\
  Engel, H.A, Emmanuel I. Rashba, E. I.and Bertrand I. Halperin, B. I. ``Theory of Spin Hall Effects in Semiconductors'', cond-mat/0603306.\\ 
 \noindent
 Entin-Wohlman et al 2010\\
Entin-Wohlman, O, Aharony, A. Tokura, Y. and Avishai Y.
``Spin-polarized electric currents in quantum transport through tubular two-dimensional electron gases'', Phys. Rev. B 81, 075439.\\
\reversemarginpar\marginnote{\bf F \ \ \ \ }Foldy 1950 \\
Foldy, L. L. and Wouthuysen, S. A. ``On the Dirac Theory of Spin 1/2 Particles and its Non-Relativistic Limit'', Phys. Rev. 78 (1): 29-36.\\
Fr\"ohlich 1993\\
Fr\"ohlic, J and Studer, U. M. ``Gauge invariance and current algebra in non-relativistic many-body theory",
J. Math. Phys. 65, 733.\\
\noindent
Frustaglia 2004\\
Frustaglia, D. Richter, K. ``Spin interference effects in ring conductors subject to Rashba coupling'', Phys. Rev. B 69, 235310.\\
\noindent
Frustaglia 2020\\
Frustaglia D. Nitta J.
``Geometric spin phases in Aharonov--Casher interference",
Solid State Communications 311, 113864. \\
\noindent
\reversemarginpar\marginnote{\bf G \ \ \ \ }Gefen {\it{et al}}. 1984\\
Gefen Y. Imry Y.  and Azbel, M. Y. 
``Quantum oscillations and the Aharonov--Bohm effect for parallel resistors",
Phys. Rev. Lett. 52 (2), 129.\\
\noindent
Grbic 2007\\
Grbic, B. Leturcq, R.  Ihn, T. Ensslin, K. Reuter, D.  Wieck, A.D. ``Aharonov--Bohm oscillations in the presence of strong spin-orbit interactions", Phys. Rev. Lett. 99, 176803.\\
\reversemarginpar\marginnote{\bf H \ \ \ \ } Hagen 1990a\\
Hagen, C. R. 
``Aharonov--Bohm scattering of particles with spin'',
 Phys. Rev. Lett. 64(5), 503. \\
\noindent
Hagen 1990b\\
Hegen C. R. ``Exact equivalence of spin-1/2 Aharonov--Bohm and Aharonov--Casher effects'',
Phys. Rev. Lett. 64, 2347.\\
 \noindent
 Hohensee {\it{et al}}. 2012\\
 Hohensee, M. A. Estey, B. Hamilton, P.  Zeilinger, A. Mueller, H. ``Force-Free Gravitational Redshift: Proposed Gravitational Aharonov--Bohm Experiment'', Phys. Rev. Lett. 108 (23): 230404. \\
  \noindent
  Heras, 2022\\
  Heras R.   ``The Aharonov–Bohm effect in a closed flux line",  Eur. Phys. J. Plus, 137, 641. \\
 \noindent 
  Hemanta {\it et. al}. 2023 \\
  Hemanta, K. K, Biswas, S.   Ofek, N. Umansky, V.  and Heiblum, Moty,  ``Anyonic interference and braiding phase in a Mach-Zehnder interferometerAnyonic interference and braiding phase in a Mach-Zehnder interferometer",  Nature Physics, January, 2023, https://doi.org/10.1038/s41567-022-01899-z. \\
\noindent
\reversemarginpar\marginnote{\bf I \ \ \ \ }Imry and Webb 1986\\
Imry, Y; Webb, R. A. ``Quantum Interference and the Aharonov-Bohm Effect'', Scientific American 260 (4): 56-62.\\
\noindent
Imry 1986\\
Imry Y. ``Active Transmission Channels and Universal Conductance Fluctuations",
Europhysics Lett, 1, 249. 
\noindent
Imry 1995\\
Imry Y. 
``Coherent Propagation of Two Interacting Particles in a Random Potential'',
Europhys. Lett. 30, 405. \\
\noindent
Imry 1997\\
Imry Y. {\it{Introduction to mesoscopic physics}}, Oxford University Press.
page 110 Fig. 5.4\\
\noindent
\reversemarginpar\marginnote{\bf K \ \ \ \ } Kane 1957\\
 Kane E. O. ``Band Structure of Indium Antimonide'', Journal of Physics and Chemistry of Solids. 1: 249. \\
\noindent
K\"oenig {\it{et al}}. 2006\\
K\"oenig, M. Tschetschetkin A. Hankiewicz, E. M. Sinova Jairo,  Hock V. Daumer V.  {\it{et al}}. ``Direct Observation of the Aharonov-Casher Phase'',
Phys. Rev. Lett. 96, 076804.\\
\noindent
\reversemarginpar\marginnote{\bf L \ \ \ \ }Landauer 1957\\
Landauer, R. ``Spatial Variation of Currents and Fields Due to Localized Scatterers in Metallic Conduction'', IBM J. Res. Dev. 1, 223.\\
\noindent
 Lee and Stone 1985\\
 Lee P. A. and  Stone A. D.
  ``Universal Conductance Fluctuations in Metals'',
Phys. Rev. Lett. 55, 1622.\\
 \noindent
 Levy {\it{et al}}. 1990\\
Levy, L. P., Dolan, G., Dunsmuir, J. and Bouchiat, H. 
``Magnetization of mesoscopic copper rings: Evidence for persistent currents'', 
Phys. Rev. Lett. {\bf 64}, 2074.\\
\noindent
Loss {\it{et al}}. 1990\\
Loss, D. Goldbart, P.  Balatsky, A.V. ``Berry's phase and persistent charge and spin currents in textured mesoscopic rings'', Phys. Rev. Lett. {\bf 65}, 1655.\\
\noindent
Li {\it{et al}}. 2022\\
Li, H. Dong, Z. Longhi, S. Liang, Q. Xie, D. and Yan, B.
``Aharonov-Bohm Caging and Inverse Anderson Transition in Ultracold Atoms'' Phys. Rev. Lett. 129, 220403.\\
\noindent
\reversemarginpar\marginnote{\bf M \ \ \ \ }Meir {\it{et al}}. 1989\\
Meir, Y.  Gefen, Y.  Entin-Wohlman, O.
``Universal effects of spin-orbit scattering in mesoscopic systems'',
Phys. Rev. Lett. {\bf 63} (7), 798.\\
\noindent
Meir et al 1990\\
Meir, Y. Entin-Wohlman, O. Gefen, Y.
``Magnetic-field and spin-orbit interaction in restricted geometries: Solvable models'', 
Phys. Rev. B 42 (13), 8351\\
\noindent
Mathur and Stone 1992\\
Mathur, H. Stone,  A.D. ``Quantum transport and the electronic Aharonov--Casher effect'', 
Phys. Rev. Lett. {\bf 68} 2964.\\
\noindent
Morpurgo 1998\\
Morpurgo, A.F., Heida, J.P., Klapwijk, T.M.,  van Wees, B.J., Borghs, G.  ``Ensemble- average spectrum of Aharonov--Bohm conductance oscillations: evidence for spin- orbit-induced Berry's phase'', 
Phys. Rev. Lett. {\bf 80}, 1050.\\
\noindent
Meijer 2002\\
Meijer, F.E. Morpurgo, A.F. Klapwijk, T.M. 
``One dimensional ring in the presence of Rashba spin-orbit interaction: derivation of the correct  Hamiltonian'', 
Phys. Rev. B {\bf 66}, 033107.\\
\noindent
Meijer 2005\\
Meijer, F.E. Morpurgo, A.F.  Klapwijk, T.M. Nitta, J. 
``Universal spin-induced time reversal symmetry breaking in two-dimensional electron gases with Rashba spin-
orbit interaction'', 
Phys. Rev. Lett. {\bf 95}, 186805.\\
\noindent
\reversemarginpar\marginnote{\bf N \ \ \ \ }Nitta 1997\\
Nitta, J.  Akazaki, T.  Takayanagi, H.  Enoki, T. ``Gate control of spin-orbit interaction in an inverted InGaAs/InAlAs heterostructure'', 
Phys. Rev. Lett. {\bf 78}, 1335.\\
\noindent
Nitta, J. Meijer,  F.E. and Takayanagi, H. ``Spin-interference device'', 
Appl. Phys. Lett. {\bf 75}, 695.\\
\noindent
Niu 2006\\
Niu Q. ``On a Proper Definition of Spin Current'', 
Phys. Rev. Lett. {\bf 96}, 076604. \\
\noindent
Nagasawa 2012\\
 Nagasawa, F. Takagi,  J. Kunihashi, Y.  Kohda, M.   Nitta, J. ``Experimental demonstration of spin geometric phase: radius dependence of time- reversal Aharonov--Casher oscillations'', 
 Phys. Rev. Lett. {\bf 108}, 086801.\\
 \noindent
 Nagasawa 2013\\
  Nagasawa, F.  Frustaglia, D.  Saarikoski, H. Richter, K.  Nitta, J. 
  ``Control of the spin geometric phase in semiconductor quantum rings'', 
  Nat. Commun. {\bf 4}, 2526.\\
  \noindent
  Nakamura et. al. 2019\\
Nakamura, J. Fallahi,  S.  Sahasrabudhe, H.   Rahman, R. Liang,  S.  Gardner, G. C. Manfra, M. J.  
``Aharonov-Bohm interference of fractional quantum Hall edge modes'', 
Nature Physics {\bf 15}, 563. \\
\noindent
\reversemarginpar\marginnote{\bf O \ \ \ \ }Onsager 1931\\
Onsager, L.  ``Reciprocal Relations in Irreversible Processes", Phys. Rev. {\bf 37} (4): 405-426.\\
 \noindent
Olariu and Popescu 1985\\
Olariu, S; Popescu, S. 
``The quantum effects of electromagnetic fluxes'', 
Rev. of Mod. Phys. {\bf 57} (2): 339. \\
\noindent
Overstreet {\it{et al}}. 2022\\
\noindent
Overstreet, C. Asenbaum, P. Curti, J. Kim, M. Kasevich, M. A.  
``Observation of a gravitational Aharonov--Bohm effect'', 
Science {\bf 375} (6577): 226-229. \\
\noindent
Osakabe 1986\\
Osakabe, N; {\it{et al}}. 1986. 
``Experimental confirmation of Aharonov-Bohm effect using a toroidal magnetic field confined by a superconductor'',
Phys. Rev. A {\bf 34} (2): 815-822.\\
\noindent
\reversemarginpar\marginnote{\bf P \ \ \ \ }Pauli 1927 \\
Pauli, W. ``Zur Quantenmechanik des magnetischen Elektrons'',
Zeitschrift f\"ur Physik {\bf 43}: 601-623.\\
\noindent
Peshkin and Tonomura 1989 \\
Peshkin M., Tonomura A., {\it{The Aharonov--Bohm effect}}, Springer-Verlag
Berlin Heidelberg New York London Paris Tokyo Hong Kong. \\
\noindent
Popescu 2010\\
Popescu S,  ``Dynamical quantum non-locality", Nature Physics {\bf 6} (3): 151-153.\\
\noindent
Pearle and Rizzi 2017\\
Pearle, P.  and Rizzi, A. ``Quantum-mechanical inclusion of the source in the Aharonov--Bohm effect", Phys. Rev. A. 95 (5): 052123. arXiv:1507.00068. \\
 \reversemarginpar\marginnote{\bf Q \ \ \ \ } Qian and Su 1994\\
 Qian, T. Z. and Su, Z.-B. ``Spin-orbit interaction and Aharonov-Anandan phase in mesoscopic rings", Phys. Rev. Lett. {\bf 72}, 2311.\\
\noindent
\reversemarginpar\marginnote{\bf R \ \ \ \ }Roy 1980\\
Roy, S. M. ``Condition for Nonexistence of Aharonov--Bohm Effect",
Phys. Rev. Lett. {\bf 44} (3): 11-14.\\
\noindent
Rashba 2006\\
 Rashba, E. I. ``Spin-orbit coupling and spin transport'', Physica E {\bf 34}, 31.\\
 \noindent
 Recher {\it{et al}}. 2007\\
 Recher, P. Trauzettel,  B.  Rycerz, A.  Blanter, Ya. M.  Beenakker, C. W. J. and Morpurgo, A. F.  ``Aharonov--Bohm effect and broken valley degeneracy in graphene rings", 
Phys. Rev. B {\bf 76}, 235404. \\
\noindent 
Rycerz and  Beenakker 2007\\
Rycerz, A. and  Beenakker, C. W. J.
"Aharonov--Bohm effect for a valley-polarized current in graphene",
arXiv:0709.3397 \\
\noindent
Rohrlich 2009\\
Rohrlich, D.  ``Aharonov--Casher Effect'', In: Greenberger, D., Hentschel, K., Weinert, F. (eds) Compendium of Quantum Physics. Springer, Berlin, Heidelberg.\\
\noindent
Rohrlich 2010\\
Rohrlich D. ``Duality in the Aharonov--Casher and Aharonov-Bohm effects",
Journal of Physics A Mathematical and Theoretical {\bf 43}(35):354028. \\
\noindent
Richter 2012\\
Richter, K.  ``Viewpoint: The ABC of Aharonov effects", Physics, 5, 22.\\
  \noindent
  Ronen et al. 2021\\
  Ronen Y., Werkmeister T., Najafabadi, D. H Pierce A. T., Anderson L. E. Jae Y.,  Si S.,  Lee U., Lee Y. H., Johnson B., Watanabe K., Takashi T., Yacoby A., and Kim  Ph.  ``Aharonov-Bohm effect in graphene-based Fabry--P\'erot quantum Hall interferometers", Nature Nanotechnology {\bf 16}, 563.\\
 \noindent
  \reversemarginpar\marginnote{\bf S \ \ \ \ }Shapiro 1983\\
  Shapiro B.
``Quantum Conduction on a Cayley Tree", 
Phys. Rev. Lett. {\bf 50}, 747.\\
  \noindent
  Stone 1985\\
 Stone, A. D. ``Magnetoresistance Fluctuations in Mesoscopic Wires and Rings",
Phys. Rev. Lett. {\bf 54}, 2692.\\
\noindent
Schwarzschild, 1986\\ 
 Schwarzschild, B. ``Currents in Normal-Metal Rings Exhibit Aharonov-Bohm Effect", Physics Today {\bf 39} (1): 17-20.\\
  \noindent
 Stern et. al. 1990\\
 Stern A. Aharonov Y. and Imry Y. ``Phase uncertainty and loss of interference: A general picture'', Phys. Rev. A {\bf 41}, 3436. \\
 \noindent
 Stern 1992\\
 Stern A. ``Berry's phase, motive forces, and mesoscopic conductivity", Phys. Rev. Lett. 68, 1022.\\
 \noindent
 Bachtold et al 1999\\
Bachtold A., Strunk C., Salvetat J-P , Bonard J-M, L. Nussbaumer F., T., Sch\"onenberger C.,
``Aharonov-Bohm oscillations in carbon nanotubes", Nature {\bf 397} (6721): 673. \\
  \noindent
 Sun and Xie 2005\\
Sun, Q. F. and Xie, X. C. 
``Definition of the spin current: The angular spin current and its physical consequences",
Phys. Rev. B {\bf 72}, 245305. \\
 \noindent
 Sjoqvist 2014\\
  Sjoqvist, E. ``Locality and topology in the molecular Aharonov-Bohm effect". Phys. Rev. Lett. 89 (21): 210401.  \\
  \noindent
  Saarikoski 2015\\
  Saarikoski,  H. V\'azquez-Lozano, J.E.  Baltan\'as, 
  J.P.  Nagasawa, F. Nitta, J. Frustaglia, D. ``Topological transitions in spin interferometers", Phys. Rev. B 91 (R), 241406.\\
 \noindent
 Shekhter et al 2022\\
 Shekhter, R. I. Entin-Wohlman, O. Jonson, M. and Aharony, A.
``Magnetoconductance Anisotropies and Aharonov--Casher Phases'',
Phys. Rev. Lett. {\bf 129}, 037704.\\
 \noindent
 Shech, 2022\\
 Shech,  E. ``Scientific understanding in the Aharonov-Bohm effect'',
  Theoria. {\bf 88}, 943.\\
\noindent
\reversemarginpar\marginnote{\bf T \ \ \ \ } Tonomura 2006\\
Tonomura A. ``The Aharonov-Bohm effect and its application to electron phase  microscopy",  Proc. Jpn. Acad., Ser. B 82. \\
\noindent
\reversemarginpar\marginnote{\bf V \ \ \ \ }van Oudenaarden et al 1998\\
van Oudenaarden, A. Devoret, M. H.  Nazarov, Yu. V. Mooij, J. E. ``Magneto-electric Aharonov-Bohm effect in metal rings", Nature. 391 (6669): 768. \\
\noindent
Vidal {\it{et al}}. 1998\\
Vidal, J. Mosseri, R.  and Dou\c cot, B. ``Aharonov-Bohm Cages in Two-Dimensional Structures'',
Phys. Rev. Lett. 81,  5888.\\
\noindent
 Vaidman 2000\\
  Vaidman L. ``Torque and force on a magnetic dipole", Am. J. Phys. 58, 978-983.\\
  \noindent Vaidman 2008\\
  Vaidman, L. ``Many-worlds interpretation of quantum mechanics",
Stanford Encyclopedia of Philosophy.\\
\noindent
Vaidman 2011\\
Vaidman, L.
``On the role of potentials in the Aharonov--Bohm effect", Phys. Rev. A86, 040101. \\
\noindent
  Vaidman, 2013\\
  Vaidman L. ``Paradoxes of the Aharonov--Bohm and the Aharonov--Casher effects", Arxiv1301/6153, published in Yakir Aharonov 80th birthday Festschrift. \\
  \noindent
  Valagiannopoulos et al. 2018\\
Valagiannopoulos, C.  Marengo, E. A.  Dimakis,  A. G.  and  Alù, A.  ``Aharonov–Bohm-inspired tomographic imaging via compressive sensing'', IET Microwaves, Antennas Propagation, 12, 1890. \\
  \noindent
  \reversemarginpar\marginnote{\bf W \ \ \ \ }Wu and Yang 1975\\
  Wu, T. T. and Yang, C. N. ``Concept of non-integrable phase factors and global formulation of gauge fields" Phys. Rev. D12, 3845-3857.\\
\noindent
  Webb {\it{et al}}. 1985\\
  \noindent
  Webb, R. A., Washburn, S; Umbach, C. P., Laibowitz, R. B.  ``Observation of $h/e$ Aharonov-Bohm Oscillations in Normal-Metal Rings". 
  Phys. Rev. Lett. 54 (25): 2696-2699.\\
  \noindent
  Willet et. al. 2013\\
	Willett, R. L.  Nayak,  C.  Shtengel, K.  Pfeiffer,  L. N.  West, K. W.  ``Magnetic-field-tuned Aharonov-Bohm oscillations and evidence for non-Abelian anyons at $\nu$ = 5/2",  Phys. Rev. Lett. 111, 186401. \\
  \noindent
 \reversemarginpar\marginnote{\bf Y \ \ \ \ }Yang and Mills 1954\\
 Yang C. N. and Mills R.  L.  ``Conservation of Isotopic Spin and Isotopic Gauge Invariance", Phys. Rev. 96, 191.\\
\noindent
Yau 2002\\
Yau, J.-B., De Poortere E.P., Shayegan, M. ``Aharonov--Bohm oscillations with spin: evidence for Berry's phase", Phys. Rev. Lett. {\bf 88}, 146801.\\
\noindent
Yerin et al. 2021\\
Yerin, Y.  Gusynin, V. P, Sharapov, S. G. and Varlamov, A. A.  ``Genesis and fading away of persistent currents in a Corbino disk geometry"
Phys. Rev. B 104, 075415.
\noindent
\reversemarginpar\marginnote{\bf Z \ \ \ \ }Zvyagin 2020)\\
Zvyagin A. A. ``Persistent currents in the two-chain correlated electron model",
Low Temperature Physics 46, 507. 
\end{widetext}
\end{document}